\documentclass{ccjnl}
\usepackage{lipsum,amsmath}
\usepackage{cuted}
\usepackage{bm}
\usepackage{multirow}
\usepackage{mathrsfs}
\usepackage{lipsum}
\usepackage[bottom]{footmisc}
\usepackage{color, soul, framed}
\newtheorem{criterion}{\bf{Criterion}}
\graphicspath{{figures/}}

\title{A Golden Decade of Polar Codes: From Basic Principle to 5G Applications}
\author{Kai Niu\inst{1,3,*}, Ping Zhang\inst{2}, Jincheng Dai\inst{1}, Zhongwei Si\inst{1}, Chao Dong\inst{1}\corinfo{niukai@bupt.edu.cn}}
\receiveddate{}
\reviseddate{}
\Editor{}

\address[1]{The Key Laboratory of Universal Wireless Communications, Ministry of Education, Beijing University of Posts and Telecommunications, Beijing 100876, China}
\address[2]{The State Key Laboratory of Networking and Switching Technology, Beijing University of Posts and Telecommunications, Beijing 100876, China}
\address[3]{Department of Broadband Communication, Peng Cheng Laboratory, Shenzhen 518055, China}

\begin{document}

\maketitle

\begin{abstract}
After the pursuit of seventy years, the invention of polar codes indicates that we have found the first capacity-achieving coding with low complexity construction and decoding, which is the great breakthrough of the coding theory in the past two decades. In this survey, we retrospect the history of polar codes and summarize the advancement in the past ten years. First, the primary principle of channel polarization is investigated such that the basic construction, coding method and the classic successive cancellation (SC) decoding are reviewed. Second, in order to improve the performance of the finite code length, we introduce the guiding principle and conclude five design criteria for the construction, design and implementation of the polar code in the practical communication system based on the exemplar schemes in the literature. Especially, we explain the design principle behind the concatenated coding and rate matching of polar codes in 5G wireless system. Furthermore, the improved SC decoding algorithms, such as SC list (SCL) decoding and SC stack (SCS) decoding etc., are investigated and compared. Finally, the research prospects of polar codes for the future 6G communication system are explored, including the optimization of short polar codes, coding construction in fading channels, polar coded modulation and HARQ, and the polar coded transmission, namely polar processing. Predictably, as a new coding methodology, polar codes will shine a light on  communication theory and unveil a revolution in transmission technology.
\end{abstract}
\section{Introduction}
\label{Introduction}
Channel coding, also named as error control coding, is not only the basic theory of modern information science, but also the core technique of modern communication system.

In 1948, C. E. Shannon \cite{Shannon} discovered that there exist some channel codes that can be decoded reliably when the code rate is no more than channel capacity whereas he did not provide the constructive coding schemes. In the past seventy years, pursuit of approaching capacity with practical en/decoding complexity is a central challenge in coding theory \cite{Survey_channelcode}. In 1990s, Turbo codes \cite{Turbocode} and LDPC codes \cite{LDPCcode} were discovered to dramatically improve the error performance of digital communication and approach the channel capacity, which are a big progress of coding theory. However, there is still a non-zero gap between the achievable rate of these codes and the channel capacity.

In 2008, Ar{\i}kan \cite{Polarcode_Arikan} invented polar codes\endnote{In fact, Stole \cite{Polarcode_Stolte} also independently found the construction method of polar codes whereas the proof of polarization phenomenon and capacity-achieving should be first credited to Ar{\i}kan.} to achieve the capacity of the symmetric channels, which is a great theoretic breakthrough. Polar codes are coined by the concept of channel polarization. That is to say, by using the channel combining and splitting, a group of $N$ identical binary-input discrete memoryless channels (B-DMC) can be transformed into a group of polarized channels, some of which become noiseless and the capacity tends to one (good channels) and others get noisy and the capacity goes to zero (bad channels). When the channel number, that is code length, tends to infinity, the ratio of the number of good channels to that of total channels will approach the capacity of original channel. Unlike the traditional channel codes, such as Turbo/LDPC codes, polar codes introduces a novel idea for the coding design.

\begin{figure}[htbp]
  \centering{\includegraphics[scale=0.52]{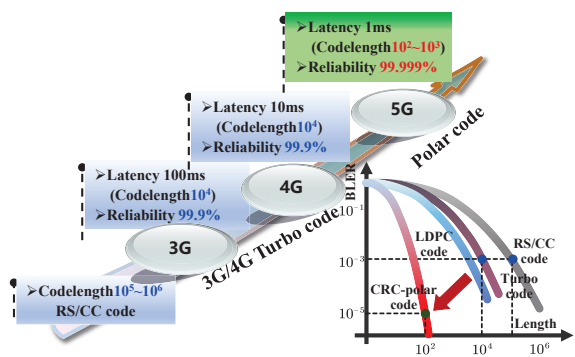}}
  \caption{Roadmap of channel coding in wireless communication systems.}\label{Fig1_Roadmap}
\end{figure}

For the practical application, channel coding is the key technology in the reliable transmission, especially in wireless communications. Figure \ref{Fig1_Roadmap} shows the roadmap of channel codes application in 3G$\sim$5G wireless systems. It follows that low-latency and high reliable transmission are the main trend of the wireless communication in the past twenty years. Low-latency requires the short code length and high reliability demands a low error rate. In pre3G wireless system, Reed-Solomon and convolutional concatenate (RS/CC) codes are applied to achieve high reliability ($99.9\%$) with a very long code length ($10^5\sim10^6$). Then in 3G and 4G systems, turbo codes are utilized to approach the same performance with a long length ($10^4$) and a low latency in terms of $10ms\sim100ms$. The 5G systems propose more rigor requirements for the transmission latency ($1ms$) and reliability ($99.999\%$) and tubo codes cannot meet this severe challenging. Due to the capacity-approaching feature, it seems that the classic polar codes can be used as the candidate of 5G channel coding.

However, the theoretic advantage of polar codes can not be transformed into the superiority of practice application. The error performance of the classic polar codes is not satisfied in the case of the finite code length. In order to improve the performance of polar codes, in the past decade, many advanced polar coding schemes and high performance decoding algorithms have been proposed. Specifically, cyclic redundancy check (CRC) concatenated polar codes and CRC aided decoding algorithms demonstrate the outstanding error performance in the short to medium code length \cite{Survey_Niu}\cite{CASCL_Niu}\cite{ASCL_Li} and outperform the performance of Turbo/LDPC codes. One highlighted merit of CRC-polar code is that it has no error floor due to the algebraic coding structure \cite{Polarcode_books}. On the contrary, both turbo and LDPC codes demonstrate the error floor effect at the high signal-to-noise ratio (SNR) region. Since the CRC-polar codes fulfill the requirement of high-reliability transmission, polar codes have been accepted as the coding standard in 5G wireless communication system \cite{5GNR_38212}.

In the previous survey papers \cite{Survey_Niu}\cite{Survey_Babar}, the basic principle of polar coding were briefly introduced. In 2019, IEEE communication society published the best readings of polar coding online \cite{Bestreadings}. In this best readings, the polar coding theory, the construction and decoding of practical polar codes, as well as practical implementations are summarized and the exemplary works are selected from the literature. The interesting reader can also refer to the book of polar codes \cite{Polarcode_books} for the further understanding.

In this paper, we retrospect the history of polar codes from birth to the present so as to review the development of this new direction. The remainder of the paper is organized as follows. Section \ref{section_II} presents the basic principle of polar codes, including channel polarization, encoding and successive cancellation decoding. Then Section \ref{section_III} describes the design criteria of polar codes with finite code length. We explain the design concerns behind the 5G polar codes. In Section \ref{section_IV}, the construction algorithms, the encoding structure of concatenated polar codes and rate compatible methods are discussed from the view point of practice application. On the other hand, we address the improved decoding algorithms of polar codes in Section \ref{section_V}, including successive cancellation list (SCL) decoding, successive cancellation stack (SCS) decoding and CRC-aided SCL/SCS decoding etc. Simulation results show the superiority of polar codes compared to Turbo/LDPC codes. Then the future direction of polar coded technique is discussed in Section \ref{section_VI}. Finally, Section \ref{section_VII} concludes the paper.

\section{Basic Principle of Polar Codes}\label{section_II}

In this section, we first explain the channel polarization, that is, the core concept of polar codes. Then, the encoding and construction of polar codes and the classic SC decoding algorithm are presented.
\subsection{Channel Polarization}
As stated in \cite{Polarcode_Arikan}, channel polarization is an interesting phenomenon between the source block and the received sequence which can be explained by using the chain rule of the mutual information. Briefly, this phenomenon can be recursively implemented by transforming multiple independent uses of a given B-DMC into a set of successive uses of synthesized binary input channels.

\begin{figure}[htbp]
  \centering{\includegraphics[scale=0.58]{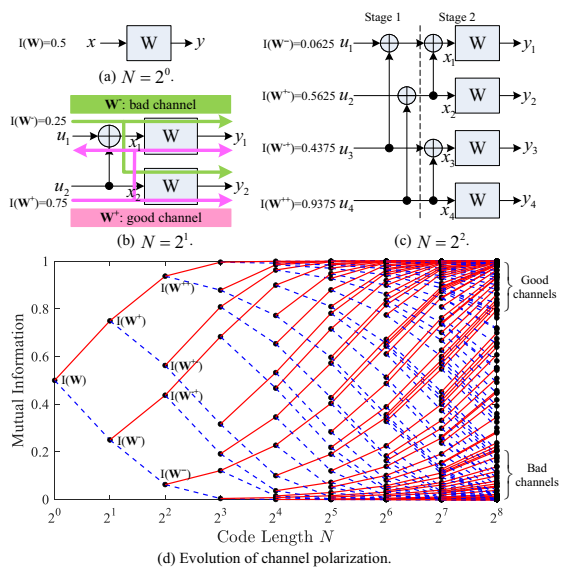}}
  \caption{Channel polarization scheme.}\label{Fig2_channel_polarization}
\end{figure}
Figure \ref{Fig2_channel_polarization} illustrates the process of channel polarization. Given one binary erasure channel (BEC) $W$ with the input bit $x$ and the output signal $y$ in Figure 2(a), the erasure probability of this BEC is 0.5 and the corresponding capacity is $I(W)=0.5$. By using one modulo-2 operation between the two independent BECs, i.e. $x_1=u_1\oplus u_2$, as shown in Figure 2(b), an equivalent compound channel can be obtained which has two input bits $u_1,u_2$ and two output bits $y_1,y_2$, as well as the associated capacity is $I(u_1,u_2;y_1,y_2)$. Furthermore, by applying the chain rule of mutual information, this compound channel can be decomposed into two synthesized channels: channel $W^{-}$  (indicated by the green line with the input bit $u_1$ and the output signals $y_1$ and $y_2$) and channel $W^{+}$ (indicated by the pink line with the input bit $u_2$  and the output signals $y_1$, $y_2$, and $u_1$). The mutual information relationship in this process can be expressed as $I(u_1,u_2;y_1,y_2)=I(u_1;y_1,y_2)+I(u_2;y_1,y_2|u_1)=I(W^{-})+I(W^{+})$.

Such operation is named as the single-step (or two-channel) polarization, which means two independent BECs with the same reliability are transformed into two polarized channels and the sum capacity of two channels is unchangeable, i.e., $I(u_1,u_2;y_1,y_2)=2I(W)$. In \cite{Polarcode_Arikan}, Ar{\i}kan proved that the bad channel $W^{-}$ has a smaller capacity than the given BEC $W$ whereas the good channel $W^{+}$ has a larger capacity, i.e., $I(W^{-})<I(W)<I(W^{+})$. Specifically, in this example, given the capacity of BEC $I(W)=0.5$, the capacities of two polarized channels are $I(W^{-})=0.25$ and $I(W^{+})=0.75$, respectively.

The single-step polarization transform can be extended to four independent BECs, as shown in Figure 2(c). Two good channels $W^{+}$ can be further transformed into two polarized channels $W^{+-}$ and $W^{++}$. Similarly, two bad channels $W^{-}$ can also be converted into two channels $W^{--}$ and $W^{-+}$. Obviously, the polarization effect is further strengthened.

In this way, the polarization transform can be recursively performed over $N=2^n$ independent uses of a given BEC. As shown in Figure 2(d), when the code length $N$ is increased from $2^0$ to $2^8$, the polarization effect demonstrates more and more prominent, i.e., the capacities of all the polarized channels tend to either 1 (good/noiseless channels marked by pink lines) or 0 (bad/noisy channels marked by dashed blue lines) except a vanishing fraction.

Theoretically, by using of martingale property, Ar{\i}kan proved the stochastic convergence behaviour of symmetric capacities of polarized channels in \cite{Polarcode_Arikan}. That is to say, given a fixed constant $\delta\in(0,1)$, as the number of channels (code length $N$) becomes infinity, the proportion of the noiseless channels is exactly equal to the symmetric capacity of the original B-DMC $W$. This limitation can be formally expressed as follows,
\begin{equation}
\lim_{N\to \infty}\frac{\left|\left\{i:I\left(W_N^{(i)}\right)>1-\delta\right\}\right|}{N}=I(W).
\end{equation}

\begin{remark}
Channel polarization is a new methodology for the channel code design. We can assign the information bits on the noiseless channels and the fixed bits on the noisy channels so as to construct the polar codes. It is well known that the joint asymptotic equipartition property (AEP) plays a central role in the proof of Channel Coding Theorem \cite{Book_Cover}. As pointed out by Niu \emph{et al.} in \cite{Survey_Niu}, channel polarization can be regarded as an analog of joint AEP. For the noiseless channels, the transmitted codeword and the received sequence forms a jointly typical mapping and about $2^{NI(W)}$ codewords can be reliably transmitted over these channels. On the other hand, the jointly typical probability between a random vector and the received sequence is approximately $2^{-NI(W)}$ tending to zero with the increasing of code length. So we can conclude that channel polarization provides a constructive proof for the channel capacity achieving.
\end{remark}

\subsection{Basic Construction and Encoding}
In order to construct polar codes, we should evaluate the reliability of each polarized channel in the channel polarization and select the high reliability channels to carry the information bits. Initially, Ar{\i}kan proposed a construction based on the Bhattacharyya parameter. Given a B-DMC channel $W:\mathcal{X}\to\mathcal{Y}$ with the input bit set $\mathcal{X}=\{0,1\}$, the output alphabet $\mathcal{Y}$ and the transition probabilities $W(y|x),x\in\mathcal{X},y\in\mathcal{Y}$, the corresponding Bhattacharyya parameter is defined as $Z(W)=\sum_{y\in \mathcal{Y}}\sqrt{W(y|0)W(y|1)}$.

Specifically, suppose a BEC $W$ with the Bhattacharyya parameters $I\left(W_1^{(1)}\right)=\epsilon$ is given and the Bhattacharyya parameters of polarized channels $Z\left(W_{N/2}^{(i)}\right), i=1,2,...,N/2$ have been obtained, the Bhattacharyya parameters of $N$ channels can be easily recursively calculated as follows
\begin{equation}
\left\{ \begin{aligned}
Z\left( {W_N^{\left( {2i - 1} \right)}} \right) &= 2Z\left( {W_{N/2}^{\left( i \right)}} \right) - {Z^2}\left( {W_{N/2}^{\left( i \right)}} \right),\\
Z\left( {W_N^{\left( {2i} \right)}} \right) &= {Z^2}\left( {W_{N/2}^{\left( i \right)}} \right).
\end{aligned}  \right..
\end{equation}

Thus we can sort the Bhattacharyya parameters $Z(W_{N/2}^{(i)})$ and select a channel index set with high reliability as the information set $\mathcal{A}$. Although this construction is a low complexity method by $O(N\log_2 N)$, it is only accurate for the coding construction in a BEC. In other B-DMCs such as the binary symmetric channel (BSC) and the binary input additive white Gaussian noise (BI-AWGN) channel, exact calculation  needs highly complicated Monte Carlo integration, while approximate iterative calculations result in the performance loss. Therefore, the Bhattacharyya parameter based method is a basic reference for the construction of polar codes.

Now we explain the encoding of polar codes. Given the code length $N$, the information length $K$ and code rate $R=K/N$, the indices set of polarized channels can be divided into two subsets: one set $\mathcal{A}$, called information set, which carries information bits and the other complement set $\mathcal{A}^c$, which is assigned the fixed binary sequence consisting of frozen bits. A message block consisting of $K=|\mathcal{A}|$ bits is transmitted over the most reliable channels $W_N^{(i)}$ with indices $i\in \mathcal{A}$, while other channels are assigned the frozen bits. Therefore, a binary source block $u_1^N$ consisting of $K$ information bits and $N-K$ frozen bits can be encoded into a codeword $x_1^N$ by
\begin{equation}\label{equation3}
x_1^N=u_1^N{\bf{G}}_N,
\end{equation}
where ${{\bf{G}}_N}={{\bf{B}}_N}{\bf{F}}_N$ is the $N$-dimensional generator matrix, ${{\bf{B}}_N}$ is the bit-reversal permutation matrix, ${\bf{F}}_N={\bf{F}}_2^{\otimes n}$ denotes the channel transformation matrix, ${{\bf{F}}_2} = \left[ { \begin{smallmatrix} 1 & 0 \\ 1 &  1 \end{smallmatrix} } \right]$ is a $2\times2$ kernel matrix and ``$^{\otimes n}$'' denotes the $n$-th Kronecker product.

\begin{figure}[htbp]
  \centering{\includegraphics[scale=0.8]{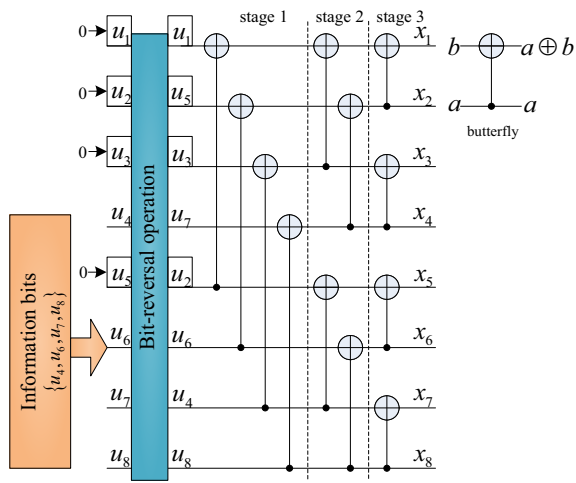}}
  \caption{An example of polar coding for $N=8,K=4$.}\label{Fig3_Polar_Coding_example}
\end{figure}

Figure \ref{Fig3_Polar_Coding_example} shows an encoder example of polar code with $N=8$, $K=4$, and $R=1/2$. According to the reliability order, the information set is $\mathcal{A}=\{4,6,7,8\}$ and the frozen set is $\mathcal{A}^c=\{1,2,3,5\}$. So the information bits $\{u_4,u_6,u_7,u_8\}$ are assigned to the polarized channels in set $\mathcal{A}$ while the frozen bits are deployed in set $\mathcal{A}^c$. Intuitively, each row of the generation matrix is associated with a polarized channel.

Furthermore, one butterfly unit is also depicted in Figure \ref{Fig3_Polar_Coding_example}, which transform two input bits $(a,b)$ into two output bits $(a\oplus b,a)$. This operation is designated by the kernel matrix $\mathbf{F}_2$ and related to the 2-channel polarization. For this example with the code length $N=8$, the polar encoder includes one bit-reversion and three stages of butterfly operations. In each stage, there are four butterfly units.

Generally, given the code length $N=2^n$, the polar encoder contains $n=\log_2 N$ stages and each stage has $N/2$ butterfly units. Thus, the encoding complexity of polar codes is $O(N\log_2 N)$. Since the bit-reversal matrix $\mathbf{B}_N$ is a symmetric and permutated matrix, it satisfies $\mathbf{B}_N^{T}=\mathbf{B}_N$ and $\mathbf{B}_N^{-1}=\mathbf{B}_N$. So the generator matrix $\mathbf{G}_N$ has two equivalent expressions and the polar encoding can be written as two forms,
\begin{equation}\label{reversal_encoder}
x_1^N=u_1^N\mathbf{G}_N=u_1^N\mathbf{B}_N\mathbf{F}_2^{\otimes n},
\end{equation}
and
\begin{equation}\label{natural_encoder}
x_1^N=u_1^N\mathbf{G}_N=u_1^N\mathbf{F}_2^{\otimes n}.
\end{equation}
The polar encoder implemented by \eqref{reversal_encoder} is named as the bit-reversal order encoder and that by \eqref{natural_encoder} is the natural order encoder.

\subsection{Successive Cancellation Decoding}
Successive cancellation (SC) decoding, proposed by Ar{\i}kan \cite{Polarcode_Arikan}, is the basic decoding algorithm of polar codes. Essentially, SC decoding is a greedy algorithm with the bit-wise soft-message calculation and hard-message decision. Hence, this decoding can be regarded as a soft/hard message passing algorithm over the trellis or code tree of polar codes.

\begin{figure}[htbp]
  \centering{\includegraphics[scale=0.8]{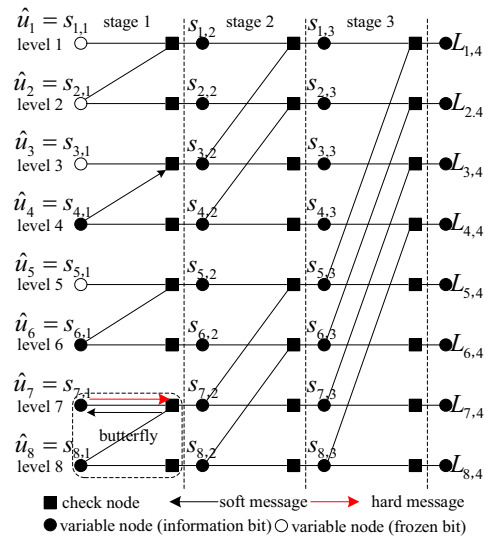}}
  \caption{Trellis of the $(8,4)$ polar code.}\label{Fig4_SC_decoder_structure}
\end{figure}

The trellis of polar codes is a regular structure with $n$ stages and $N$ levels. Each stage includes $N/2$ butterfly units and each one includes a pair of check and variable node. Figure \ref{Fig4_SC_decoder_structure} shows an example of trellis for the $(8,4)$ polar code. The soft and hard messages are calculated and passed over the variable nodes in the trellis, denoted by $s_{i,j}, 1\leq i\leq N, 1\leq j\leq n+1$. The corresponding soft messages are the logarithmic likelihood ratios (LLRs) denoted by $L_{i,j}=L\left(s_{i,j}\right)$. Similarly, the hard messages $B_{i,j}$ are the bits associated to the nodes $s_{i,j}$.
Specially, in the left side of the trellis, the soft messages of the variable are designated as $L_{i,1}=L\left(\hat {u}_i\right)$, where $\hat{u}_i=s_{i,1}$ are the estimated values of the source block. Similarly, in the right side of the trellis, the associated LLRs are denoted by $L_{i,n+1}=\log\frac{P(y_i|1)}{P(y_i|0)}$, where $y_i$ are the received signals.

The soft/hard message update and decision rule can be summarized as follows.
\begin{enumerate}
\item {\bf{Soft Message Updated Rule}}
\begin{equation}\label{SoftMsg}
L_{i,j}\!=\!\left\{
\begin{aligned}
&2{\rm{artanh}}\left[ \tanh \left( {\frac{L_{i,j+1}}{2}} \right) \right.\\
&\qquad \qquad \quad \left.\cdot \tanh \left( {\frac{L_{i+2^{i-1},j+1}}{2}} \right) \right],\\
&{\text{if}} \left\lfloor {\frac{i-1}{2^{j-1}}} \right\rfloor \bmod 2 = 0;\\
&\hspace{-0.2em}( {1 - 2{B_{i - {2^{j - 1}},j}}} )( {{L_{i - {2^{j - 1}},j + 1}}} ) \hspace{-0.1em}+\hspace{-0.1em} {L_{i,j + 1}},\\
&{\text{if}} \left\lfloor {\frac{i-1}{2^{j-1}}} \right\rfloor \bmod 2 = 1.
\end{aligned}
 \right.,
\end{equation}
where $i=1,2,\cdots, N$, $j=1,2,\cdots, n$, $\lfloor\cdot\rfloor$ is the floor function, $\tanh(\cdot)$ and $\rm{artanh}(\cdot)$ are the hyperbolic tangent function and the inverse function respectively.
The soft message at the check node labeled by the index $\lfloor(i-1)/2^{j-1}\rfloor \bmod 2 =0$ (the even number node) is updated by using the first formula of \eqref{SoftMsg} and the message at the variable node (the odd number node) is designated by the second equation.
\item {\bf{Hard Message Updated Rule}}
\begin{equation}\label{HardMsg}
B_{i,j + 1} = \hspace{-0.2em}\left\{ \begin{aligned}
&\hspace{-0.2em}B_{i,j} \hspace{-0.1em}\oplus \hspace{-0.1em}B_{i+{2^{j-1}},j},\left\lfloor {\frac{i-1}{2^{j-1}}} \right\rfloor \hspace{-0.3em} \bmod  2\hspace{-0.1em}= \hspace{-0.1em}0,\\
&\hspace{-0.2em}B_{i,j},{\text{otherwise}}.
\end{aligned} \right.,
\end{equation}
where $\oplus$ is the modulo-2 operation. In the even number node, the hard message is calculated by the first formula of \eqref{HardMsg} and in the odd number node, the message is referred to the second equation.
\item {\bf{Decision Rule}}

When the soft message is obtained in stage $1$, the bit decision rule is written as
\begin{equation}
\left\{
\begin{aligned}
&\hat{u}_i=0,  L_{i,1}\geq 0 {\text{  and  }} i \in \mathcal{A},\\
&\hat{u}_i=1,  L_{i,1}< 0 {\text{  and  }} i \in \mathcal{A},\\
&\hat{u}_i=0,  i\in \mathcal{A}^c.
\end{aligned}\right..
\end{equation}
Therefore, for an information bit, the bit value associated to the node is decided by using the soft message and for a frozen bit, it can be simply set to a pre-defined value, i.e., $\hat{u}_i=0$.
\end{enumerate}

The computational complexity of SC decoding is mainly determined by the soft message calculations in the butterfly units. Since the trellis consists of $N/2\log_{2}N$ butterfly units, the time complexity of the SC decoder is $O(N\log_2 N)$. Intuitively, the SC decoder needs $N\log_2N$ memory units to store the LLRs. However, this memory consumption can be reduced to $N$ units using the memory sharing mechanism.

Given an $(N,K)$ polar code with the information set $\mathcal{A}$, the block error rate (BLER) under SC decoding can be upper bounded by \cite{Polarcode_Arikan}
\begin{equation}
P_e(N,K,\mathcal{A})\leq \sum_{i\in \mathcal{A}} Z\left(W_N^{(i)}\right).
\end{equation}
When the code length $N$ goes to infinity, the asymptotic BLER trends to $o\left(2^{\sqrt{N}}\right)$, that is to say, the block error probability of polar codes will exponentially decay with the square root of the code length \cite{polar_rate}.

In a word, the theoretical advantages of polar codes can be summarized as follows.
\begin{enumerate}
\item {\bf{Channel capacity achieving}}\\
Polar codes are a class of constructive channel codes achieving the symmetric capacity of the binary-input discrete memoryless channels. It firstly reveals the coding process approaching the channel capacity and equivalently provides a constructive proof of the channel coding theorem.
\item {\bf{Low complexity en/decoding}}\\
By using the Walsh-Hardamard transform, the encoding complexity of polar codes is $O(N\log_2 N)$. Correspondingly, using SC algorithm, the decoding complexity of polar codes is also $O(N\log_2 N)$. Both the encoding and decoding have low complexity.
\item {\bf{Error floor free}}\\
Due to the algebraic structure of polar codes, the block error probability under SC decoding exponentially decreases with the square root of the code length. Such algebraic property determines the polar codes have no error floor. On the contrary, both turbo and LDPC codes have the error floor phenomenon.
\item {\bf{Channel polarization universality}}\\
Channel polarization is a universal phenomenon in the communication scenario rather than a specific technique in coding system. So it provides a novel idea and methodology to design and optimize the communication systems.
\end{enumerate}

Unfortunately, polar codes with finite length have some critical drawbacks. From the viewpoint of practical application, we list the main problems of the original polar codes should be solved.
\begin{enumerate}
\item {\bf{Channel dependent construction}}\\
The initial construction based on Bhattacharyya parameter is a channel dependent algorithm and not precise for a general B-DMC. So it is important to precisely evaluate the reliability of polarized channels for any B-DMC. Furthermore, constructing a polar code independent of the channel condition is desirable for practical application.
\item {\bf{Small minimum-Hamming distance}}\\
In the case of finite length, the minimum Hamming distance of polar codes is very small and inferior to some famous algebra codes, such as Reed-Muller (RM) codes or BCH codes. This drawback leads the performance of polar codes under maximum likelihood (ML) decoding is worse than RM/BCH codes. Hence, it is significant to improve the minimum-Hamming distance of polar codes by using algebraic coding techniques.
\item {\bf{Coding length constraint}}\\
Due to the structure constraint of generator matrix of polar codes, the code length is restricted to the power of two, such as $N=2^{10}=1024$. However, in the data transmission, the code length is required to be arbitrary and flexible, i.e., $N=800,1000$, etc. Hence, this constraint limits the application of polar codes in communication systems. We should design a rate compatible polar code to fulfill the practice requirements.
\item {\bf{Poor error performance of SC decoding}}\\
The SC decoding is a suboptimal algorithm since the calculation and decision bit-by-bit may result in the error propagation. So the error performance of SC decoding is worse than that of turbo/LDPC codes. This is a big barrier for the polar code applying in communication systems. We should design a more powerful decoding algorithm to improve the error performance of polar codes.
\end{enumerate}
In the next sections, we will explain the design rules of finite length polar codes, especially for the design principle behind the 5G polar codes.

\section{Design Criteria of Finite Length Polar Codes}
\label{section_III}
\begin{figure*}[htbp]
  \centering{\includegraphics[scale=0.8]{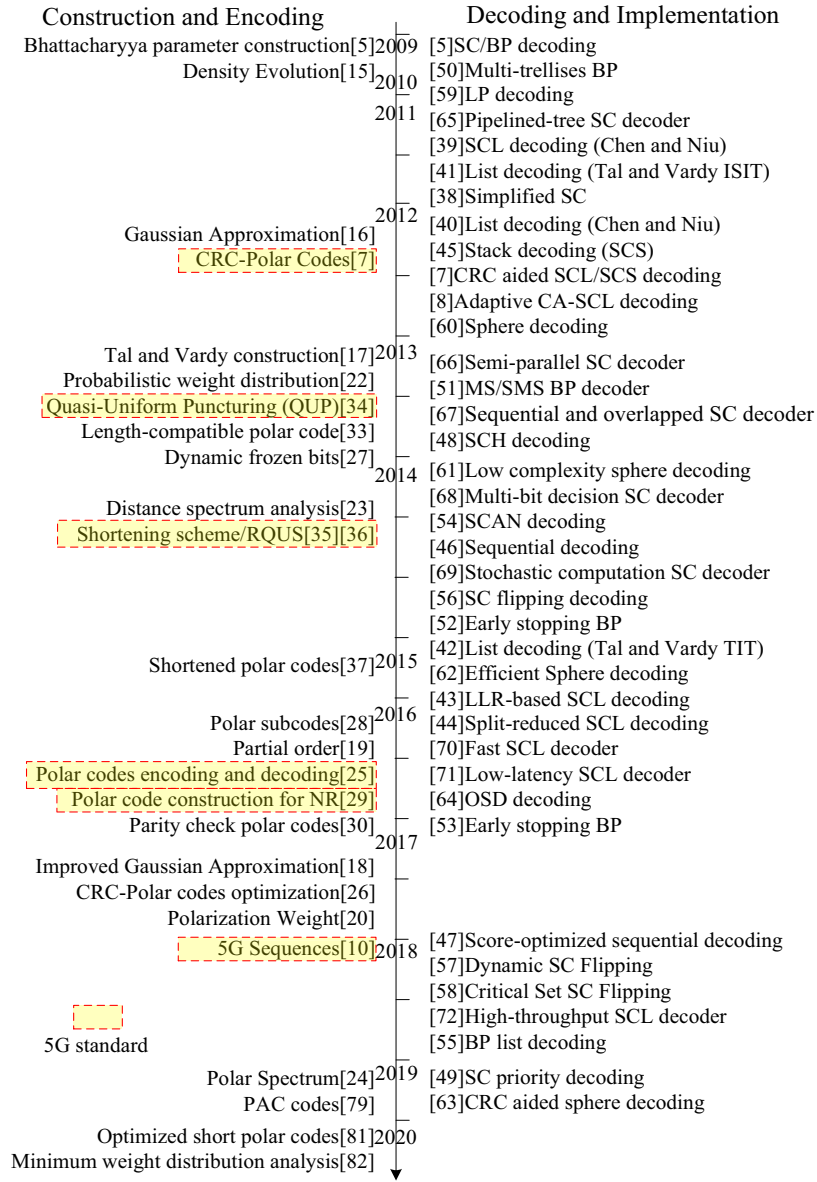}}
  \caption{Major contributions to the polar coding.}\label{Fig5_Timeline_Polar_Code}
\end{figure*}

\begin{figure*}[htbp]
  \centering{\includegraphics[scale=0.7]{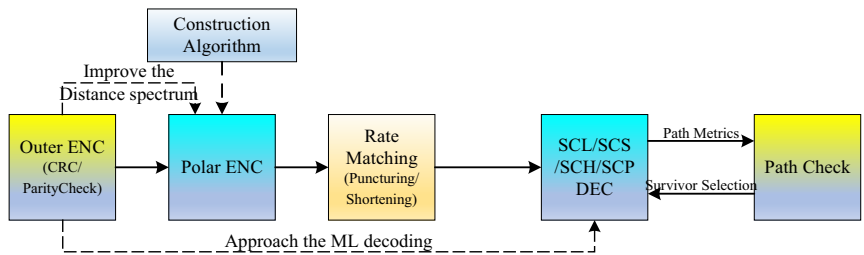}}
  \caption{The design framework of finite length polar codes.}\label{Fig6_Concatenated_Polar_Coding_Scheme}
\end{figure*}

In the past decade, many works were proposed to overcome the problems of the polar codes with finite length. Figure \ref{Fig5_Timeline_Polar_Code} shows the main contributions to the polar coding. Considering the requirements of practical communications, the performance of polar codes with finite length can be improved by jointly designing the encoding structure and decoding algorithm. Thus, we can summarize the en/decoding design framework of polar codes as shown in Figure \ref{Fig6_Concatenated_Polar_Coding_Scheme}.

The concatenated coding consists of the inner polar code and the outer code (e.g. cyclic redundancy check (CRC) code) is the main form of powerful polar codes with finite length. In this framework, the outer codes play double roles. At the side of the encoder, suitable outer codes can raise the minimum Hamming distance to improve the performance of the concatenated codes. Furthermore, we should solve the problems of construction, encoding, and rate compatibility. on the other hand, at the side of the decoder, the outer codes can be used to check the survivor paths and select the correct codeword whereby dramatically enhance the error performance. Therefore, powerful decoding algorithms and hardware implementation are the main concerns. Guided by this framework, the design criteria of polar codes with finite length can be summarized as follows.

\begin{criterion}
\bf{Polar codes should be constructed by high-precise, low-complexity and channel-independent methods.}
\end{criterion}

The existing construction methods of polar codes can be mainly divided into three categories: (1) channel dependent construction, (2) channel independent construction, (3) weight distribution based construction.

In the first category, some channel parameters, e.g., signal-to-noise ratio (SNR) in binary-input AWGN channel, are utilized to evaluate the reliability of the polarized channel based on the recursive structure of the polar coding. Recall that the Bhattacharyya parameter based construction is accurate only for the coding construction in a BEC yet approximate for other B-DMCs. Subsequently, Mori \emph{et al.} \cite{DE_Mori} designed a density evolution (DE) algorithm to track the distribution of LLR and calculate error probability under the successive cancellation (SC) decoding with a complexity of $O(N\xi \log{\xi})$. But high theoretical accuracy will need large number of samples $\xi$. Afterward, Trifonov \cite{GA_Trifonov} advocated the use of Gaussian approximation (GA) construction with a complexity of $O(N)$. Tal and Vardy \cite{Tal_Vardy} proposed an iterative algorithm to evaluate the upper/lower bound on error probability of each polarized channel with a complexity of $O(N{\mu}^2\log\mu)$, where $\mu$ is a fixed integer called the fidelity parameter. GA can approach the Tal-Vardy method with a lower complexity. Then Dai and Niu \emph{et al.} \cite{IGA_Dai} proposed an improved GA algorithm for the polar codes with the long code length to further increase the accuracy of the GA construction.

In the second category, the reliability of the polarized channel can be ordered based on some channel-independent characteristics of polar codes. This construction is desirable for the practical implementation. Sch$\rm\ddot{u}$rch \emph{et al.} \cite{PO_Schurch} introduced the concept of partial order (PO) to indicate the invariant reliable order of a part of polarized channels. Furthermore, He \emph{et al.} \cite{PW_He} proposed the polarized weight (PW) algorithm. Although PW is an empirical construction, it can amazingly achieve almost the same performance as those constructed by the GA algorithm. In fact, the polar codes adopted in 5G standard \cite{5GNR_38212} are constructed through a fixed reliability sequence of polarized channels obtained by a computer searching \cite{5Gpolar_Bioglio}.

In the third category, distance spectrum or weight distribution is considered to thoroughly interpret the behavior of polar codes and deduce the construction of polar codes. Valipour and Yousefi \cite{Weight_distribution} designed a probabilistic method to estimate the weight distribution of polar codes whereas the computational complexity is very high. Then, by using successive cancellation listing algorithm, \cite{ASCL_Li} and \cite{Distance_spectrum} proposed the methods to enumerate and calculate the distance spectrum of polar codes. Recently, Niu \emph{et al.} \cite{Polar_spectrum} introduced a new theoretical tool, named polar spectrum, to analyze and construct the polar codes. Polar spectrum provides a new insight to understand the algebraic property and has the potential value for the design of polar codes.

\begin{criterion}
\bf{Concatenated coding is a powerful coding scheme to improve the error performance of polar codes.}
\end{criterion}

In the case of finite length, increasing the minimum Hamming weight is the core idea to enhance the error performance of polar codes. According to the classic coding theory, concatenated coding is a suitable method to fulfill this aim. As shown in Figure \ref{Fig6_Concatenated_Polar_Coding_Scheme}, the compound encoder consists of the inner polar encoder and the outer encoder. In the literature, Niu and Chen \cite{CASCL_Niu} firstly proposed the cyclic redundancy check concatenated polar (CRC-polar) codes to improve the error performance of short to medium length, which is a typical example of concatenated coding. Since CRC-polar codes increase the minimum Hamming distance and weight distribution, the corresponding performance can dramatically outperform that of turbo/LDPC codes. As a high efficiency and low complexity coding scheme, CRC-polar code was cited by 5G proposal \cite{CRCPolar_Huawei} and accepted as the basic coding in 5G standard \cite{5GNR_38212}. Furthermore, Zhang \emph{et al.}\cite{CRC_opt_Zhang} investigated the optimization of CRC generator polynomials.

Trifonov and Miloslavskaya \cite{Dynamic_Trifonov}\cite{Polar_Subcode_Trifonov} proposed a dynamic-frozen coding scheme by combining the eBCH codes and polar codes whereas this scheme is not flexible for arbitrary coding configurations. Guided by the idea of concatenated coding, Huawei corporation proposed a parity check concatenated polar (PC-polar) codes \cite{PCPolar_Huawei} and the corresponding coding scheme was incorporated into 5G standard \cite{5GNR_38212}. An similar concept and initial coding were also introduced in \cite{PC_Polar_Wang}. Chen \emph{et al.} \cite{Polar_Hash} designed the hash-polar code to improve the performance of polar codes under the SCL decoding. Recently, Zhou \emph{et al.} \cite{Polar_Gene} proposed a genetic algorithm to construct the polar codes so as to obtain the performance gain for SCL decoding.

As a representative coding scheme, CRC-polar codes guide a new direction to improve the performance of polar codes. In the next section, we will explain the technical details

\begin{criterion}
\bf{Rate matching is the critical requirement for the application of polar codes.}
\end{criterion}

For the original polar code \cite{Polarcode_Arikan}, the code length $N$ is limited to the power of two, i.e., $N=2^n$. Consequently, designing good rate-compatible polar (RCP) codes, also named rate matching, to flexibly support an arbitrary code length and code rate becomes the key issue to practical application of the polar codes, which usually falls into two categories: puncturing and shortening.

For the puncturing mode, Shin \emph{et al.} \cite{LCPC_Shin} advocated the use of a reduced generator matrix for efficiently improving the error performance of the RCP codes under the successive cancellation (SC) decoding yet searching good polarizing matrices is a time-consuming process. Niu \emph{et al.} \cite{QUP_Niu} proposed an efficiently universal puncturing scheme, named quasi-uniform puncturing algorithm (QUP) and the resulting RCP codes outperform the turbo codes used in 3G/4G wireless systems. Due to the high-performance and flexibility, this scheme was cited by 5G proposal \cite{CRCPolar_Huawei} and used as the basic puncturing scheme of polar codes in 5G standard \cite{5GNR_38212}.

On the other hand, for the shortening mode, Wang \emph{et al.} \cite{Shorten_WangLiu} devised a simple method by shortening the columns with weight $1$ which is the same as the reversal quasi-uniform shortening (RQUS) algorithm proposed in \cite{RateMatch_Niu}. Such simple and high efficient scheme  is also accepted as the basic shortening scheme in 5G standard \cite{5GNR_38212}. Meanwhile, Miloslavskaya \cite{Shorten_PC} exploited the structure of polar codes and jointly optimized the shortening patterns and the source-bit assignment whereas the searching complexity is still very high and the shortening pattern is not universal.

\begin{criterion}
\bf{Designing the high-performance and low-complexity algorithms is the key issue of polar decoding.}
\end{criterion}

In the past decade, polar decoding algorithms were deeply investigated based on multiple mechanisms and viewpoints. Generally, the decoding algorithms can be mainly divided into four categories: (1) SC enhanced decoding, (2) soft output decoding, (3) SC flipping decoding, and (4) decoding for short code.

In the first category, we mainly simplify or enhance the SC decoding over the trellis or code tree. Alamdar-Yazdi and Kschischang \cite{SSC} proposed a simplified successive-cancellation (SSC) decoder to decrease the redundant calculations in SC decoding without affecting the error performance. For the improved SC decoding algorithms, the most important algorithm is the successive cancellation list (SCL) decoding, which was independently proposed by two groups, Chen-Niu \cite{SCL_Conf}\cite{SCL_JNL} and Tal-Vardy \cite{Tal_Vardy_ISIT}\cite{Tal_Vardy_TIT}. Unlike SC only decodes one path at each level, Chen and Niu \cite{SCL_Conf}\cite{SCL_JNL} realized that the greedy search of SC on the code tree can be extended to a width-first search. So a maximum of $L$ candidate paths are reserved as the survivor path list and the final decision can be selected from the list. Tal and Vardy \cite{Tal_Vardy_ISIT}\cite{Tal_Vardy_TIT} proposed the similar mechanism and designed a so-called ``lazy-copy" operation to reduce the memory overhead of path copy. Thanks to the algebraic structure of polar codes, with a small list size (e.g. $L=32$) at the medium length (e.g. $N=1024$), the BLER performance of SCL decoding can approach that of ML decoding. Meanwhile, the complexity of SCL is $O(LN\log_2 N)$. SCL decoding is a great step to improve the performance of polar codes. Then Balatsoukas-Stimming \emph{et al.} \cite{SCL_LLR} designed the SCL decoder based on LLR calculation in order to facilitate the hardware implementation. Afterward, Zhang \emph{et al.} \cite{SCL_Zhang} proposed the split reduced SCL decoder to further reduce the average decoding complexity.

On the other hand, Niu and Chen \cite{SCS_Niu} realized that polar decoding can also performed a depth-first search on the code tree and proposed the successive cancellation stack (SCS) decoding. This algorithm uses an ordered stack to store the candidate paths and dramatically decrease the computational complexity down to $O(N\log_2N)$ approaching that of SC decoding. Lately, based on the similar idea, Trifonov \emph{et al.} \cite{SCSeq_Trifonov} proposed the sequential decoding and the path metric is optimized in \cite{SCS_Trifonov} to further efficiently reduce the algorithm complexity.

Combining the principles of SCL and SCS, Chen and Niu \emph{et al.} \cite{SCH_Chen} proposed the successive cancellation hybrid (SCH) decoding so as to provide a flexible configuration when the time and space complexities are limited. Lately, Guan and Niu \emph{et al.} \cite{SCP_Guan} designed a successive cancellation priority (SCP) decoding to decrease the number of path sorting operations.
Under proper configurations, all these improved SC algorithms (SCL/SCS/SCH/SCP) can approach the performance of ML decoding but with an acceptable complexity.

To further improve the performance of polar codes, Niu and Chen \cite{CASCL_Niu} proposed the CRC-aided SCL/SCS (CA-SCL/SCS) decoding schemes. In these schemes, the SCL/SCS decoder outputs the candidate paths into a CRC detector, and the check results are utilized to detect the correct codeword. To lower the time complexity of SCL decoding brought by a large list size, Li \emph{et al.} \cite{ASCL_Li} proposed an adaptive CRC-aided SCL decoder (aCA-SCL) by gradually increasing the list size. Specifically for short to medium code length, the CA-SCL/SCS decoding can substantially improve the performance of polar codes and outperform turbo/LDPC codes. This is the key advantage of polar codes that can be adopted in 5G standard.

In the second category, we mainly focus on improving the performance and throughput of the BP decoder. Ar{\i}kan \cite{Polarcode_Arikan} first pointed out that BP algorithm can be used to decode polar codes over the trellis. The computational complexity of BP decoding is $O(I_{max} N\log_2 N)$, where $I_{max}$ is the maximum number of iterations. Then, Hussami \emph{et al.} \cite{BP_original} designed the multiple trellis BP algorithm to improve the performance of standard BP. Yuan and Parhi \cite{BP_Yuan} proposed the min-sum (MS) and scaled min-sum (SMS) algorithm to simplify the BP decoder. In addition, two early stopping criteria were investigated in \cite{BPEarly_Yuan} and \cite{BPEarly_Simsek} respectively in order to decrease the iterative number and lower the complexity of BP decoding. Furthermore, by using the soft information calculation based on SC-like scheduling, Fayyaz and Barry \cite{SCAN_Fayyaz} introduced the low-complexity soft-output decoding, named soft cancellation (SCAN), to make a tradeoff between the performance and complexity. Unfortunately, all the above algorithms are inferior to the SCL decoding. Recently, Elkelesh \emph{et al.} \cite{BPL_Elkelesh} proposed the belief propagation list (BPL) decoding, which can approach SCL at the cost of increasing the complexity.

In the third category, our primary concern is the successive cancellation flipping (SCF) decoding to approach the performance of SCL with a low complexity. Afisiadis \emph{et al.} \cite{SCF_Afisiadis} first proposed the bit-flipping method to generate multiple decision attempts. Then Chandesris \emph{et al.} \cite{SCF_Chandesris} introduced the concept of high order bit flips and designed a dynamic SCFlip decoding to keep the balance between the error performance and the average complexity. Zhang \emph{et al.} \cite{SCF_Zhang} constructed a critical-set tree search to build a progressive bit-flipping decoding to efficiently lower the average complexity. Although  these SCF decoding algorithms have low average complexity, the computational complexity in the worst case is still very high.

In the fourth category, we pay attention to the (quasi)-maximum likelihood decoding of short polar codes. Goela \emph{et al.} \cite{LP_Goela} introduced the linear programming (LP) decoding for polar codes whereas this algorithm is only suitable for the BEC channel. Kahraman and Celebi \cite{SD_Kahraman} realized the lower triangle structure of the generator matrix and introduced the sphere decoding (SD). Niu \emph{et al.} \cite{SD_Niu} simplified the standard SD based on the optimum path metrics. Lately, Guo and Fbregas \cite{SD_Guo} designed the fixed and dynamic bounds to further reduce the complexity of SD. Piao and Niu \emph{et al.} \cite{SD_Piao} considered the structure of CRC-polar codes and designed the CRC-aided SD (CA-SD) in order to achieve the performance of ML decoding. In addition, Wu \emph{et al.} \cite{OSD_Wu} proposed the ordered statistic decoding (OSD) to approximate the ML decoding of short polar codes. Although SD or OSD decoding can approach the performance of ML decoding, the computational complexity of these algorithms is very high. Therefore, these algorithms are only applied for decoding polar codes with short block length.

\begin{criterion}
\bf{Designing the high-throughput and low-latency architecture is the key issue of hardware implementation.}
\end{criterion}

For the hardware implementation, the SC and SCL decoder with high-throughput and low-latency are pursued by the practical application. Leroux \emph{et al.} proposed the pipelined-tree architecture in \cite{HardD_Leroux} and the semi-parallel architecture in \cite{Semi_Leroux} respectively to improve the throughput of SC decoder. Then Zhang and Parhi \cite{LowD_Zhang} designed the sequential and overlapped architecture to further reduce the decoding latency of SC decoder. In addition, Yuan and Parhi \cite{LowD_Yuan} introduced the multi-bit decision to improve the throughput of SC decoder. Specifically, Xu and Niu \cite{Sto_Xu}
designed an SC decoder based on stochastic computation which is a new architecture with high-throughput and low power assumption.

On the other hand, many works focus on the hardware implement of SCL decoder. Sarkis \emph{et al.} \cite{SCLD_Gross} proposed a fast list decoder based on the idea of SSC to improve the throughput. Fan \emph{et al.} \cite{SCLD_Fan} considered the path selective expansion and double threshold fast-sorting method to reduce the computation and latency of SCL decoder. Recently, Xia \emph{et al.}
\cite{SCLD_Xia} designed a high-throughput SCL decoder achieves the decoding throughput of $1.103$ Gbps with the list size $L=8$.

\begin{remark}
In 5G wireless communication system, high-reliability is the basic requirement of the data transmission. The designing of 5G polar codes should keep the balance between the high performance and the implementation complexity. Although the DE/GA/Tal-Vardy constructions have high precision, the channel mapping sequence \cite{5GNR_38212} is adopted in 5G standard due to the advantages of channel-independence and low complexity. Since the CRC-polar codes \cite{CASCL_Niu} have simple structure and excellent performance, they are utilized as the basic coding scheme in 5G standard. Meanwhile, PC-polar codes \cite{PCPolar_Huawei} are also adopted as the further supplement. Furthermore, in order to fulfill the requirement of flexible coding, the QUP \cite{QUP_Niu} and RQUS \cite{Shorten_WangLiu}\cite{RateMatch_Niu} schemes are used as the rate matching methods in 5G standard. Certainly, behind all these design concerns, the CRC aided SCL/SCS decoding \cite{CASCL_Niu} is the most important factor whereby the polar codes can be applied in 5G. By now, CA-SCL/SCS decoding has become the most popular algorithm and the standard reference in the field of polar codes.
\end{remark}

\section{Construction and Encoding of Polar Codes}
\label{section_IV}
In this section, we first investigate the efficient construction methods, such as GA and PW. Then the minimum distance and weight distribution of CRC-Polar codes are analyzed to reveal the advantage of concatenated coding. Finally, we briefly explain the basic principle of QUP and RQUS scheme. All these methods are the critical parts of the polar encoding in the case of finite length.
\subsection{Efficient Construction Methods}
Gaussian approximation (GA) construction \cite{GA_Trifonov} is suitable for the polar coding in the AWGN channel. Suppose the coded bits are modulated using binary phase shift keying (BPSK) and transmitted over the AWGN channel with noise variance $\sigma^2$. The transition probability is written as $W(y|x)=\frac{1}{\sqrt{2\pi \sigma^2}}e^{-\frac{(y-(1-2x))^2}{2\sigma^2}}$, where $x\in \{0,1\}$ and $y\in\mathbb{R}$. The LLR of each receive signal $y$ is denoted by $L(y)=\ln \frac{W(y|0)}{W(y|1)}=\frac{2y}{\sigma^2}$, which obeys Gaussian distribution, that is, $L(y)\sim \mathcal{N}\left(\frac{2}{\sigma^2},\frac{4}{\sigma^4}\right)$. Due to the symmetry of polar codes, an all-zero codeword is assumed to transmit.

In GA construction, the LLR of each polarized channel $W_N^{(i)}$ is assumed to obey a Gaussian distribution, that is, $L_N^{(i)}\sim \mathcal{N}\left(m_N^{(i)},2m_N^{(i)}\right)$, where $m_N^{(i)}$ is the LLR mean. Since the LLR mean indicates the reliability of each channel, we can trace these LLR means to select the good channels to carry the information bits.

Suppose an initial LLR mean $m_1^{(1)}=\frac{2}{\sigma^2}$ is given and the LLR means of polarized channels $Z\left(W_{N/2}^{(i)}\right), i=1,2,...,N/2$ have been calculated, the LLR means of $N$ channels can be computed recursively as follows
\begin{equation}
\left\{ \begin{aligned}
m_N^{(2i-1)}  &= \phi^{-1}\left[1-\left(1-\phi\left(m_{N/2}^{(i)}\right)\right)^2\right],\\
m_N^{(2i)}     &=2 m_{N/2}^{(i)}.
\end{aligned}  \right.,
\end{equation}
where the function $\phi(\cdot)$ is defined as
\begin{equation}
\phi(t)=\left\{
\begin{aligned}
&1\hspace{-0.2em}-\hspace{-0.2em}\frac{1}{\sqrt{4\pi t}}\int_{\mathbb{R}}\hspace{-0.3em}\tanh\left(\frac{z}{2}\right)\hspace{-0.2em}e^{-\frac{(z-t)^2}{4t}}dz, &\hspace{-0.3em}t>0,\\
&1, &\hspace{-0.3em}t=0.
\end{aligned}
\right..
\end{equation}

Obviously, the exact calculation of LLR means in check nodes involves complex integration at the cost of high computational complexity. Generally, in conventional GA construction, we can use the well known two-segment function $\varphi(t)$ to approximate $\phi(t)$,
\begin{equation}
\varphi(t)=\left\{
\begin{aligned}
&e^{-0.4527t^{0.86}+0.0218}, &0<t<10,\\
&\sqrt{\frac{\pi}{t}}e^{-\frac{t}{4}}\left(1-\frac{10}{7t}\right), &t\geq10.
\end{aligned}
\right..
\end{equation}

Since this two-segment function $\varphi(t)$ has calculation error compared to the function $\phi(t)$, the channel selection may be inaccurate when the code length becomes long. So in \cite{IGA_Dai}, this function is modified by the more precise approximations, such as the new two-segment function $\Omega_2(t)$, three-segment function $\Omega_3(t)$ and four segment function $\Omega_4(t)$. They are defined as follows,
\begin{equation}
\Omega_2(t)=\left\{
\begin{aligned}
&e^{0.012t^2-0.421t} &0<t\leq7.063,\\
&e^{-0.294t-0.317}    &t>7.063.
\end{aligned}
\right.,
\end{equation}

\begin{equation}
\Omega_3(t)=\left\{
\begin{aligned}
&e^{0.0673t^2-0.491t} &0<t\leq0.636,\\
&e^{-0.453t^{0.86}+0.022} &0.636< t \leq 9.225,\\
&e^{-0.283t-0.425} &t>9.225.
\end{aligned}
\right.,
\end{equation}
and
\begin{equation}
\Omega_4(t)=\left\{
\begin{aligned}
&e^{0.105t^2-0.499t} &0<t\leq0.191,\\
&0.998e^{0.053t^2-0.480t} &0.191<t\leq0.742,\\
&e^{-0.453t^{0.86}+0.022} &0.742<t\leq9.225,\\
&e^{-0.283t-0.425} &9.225<t.
\end{aligned}
\right..
\end{equation}
Using these modified approximation functions in the improved GA construction, the polar codes with the long code length (e.g. $N=2^{14}\sim2^{18}$) can ensure the excellent performance in BI-AWGN channels.

Another typical construction is the polarization weight (PW) metric \cite{PW_He}. Given the code length $N=2^n$, by using the so-called $\beta$ expansion, PW metric can be calculated as follows,
\begin{equation}\label{PW}
PW_N^{(i)}=\sum_{s=1}^{n}i_s\beta^{n-s},
\end{equation}
where $\beta$ is the weight factor and $\left(i_1,i_2,\cdots,i_n\right)$ is the binary expansion vector associate to the channel index $i$, that is, $i-1=\sum_{s=1}^n i_s2^{s}$. When the code length is fall in the range $N=16\sim 1024$, the weight factor is optimized as $\beta=2^{1/4}\approx1.1892$.

Obviously, by \eqref{PW}, PW is a channel-independent metric. The larger the metric $PW_N^{(i)}$, the more reliable the corresponding channel $W_N^{(i)}$. On the basis of PW, after some adjustment by using the computer simulation, the mapping sequence of polar codes \cite{5GNR_38212} is obtained in 5G standard.

\subsection{CRC Concatenated Polar Codes}
\begin{figure}[htbp]
  \centering{\includegraphics[scale=0.7]{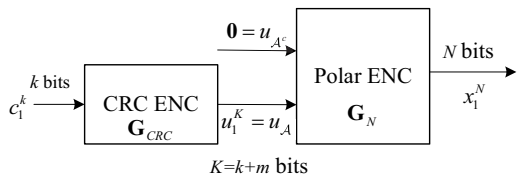}}
  \caption{CRC-polar encoder.}\label{Fig7_CRC_Polar_ENC}
\end{figure}

CRC-polar codes are the primary concatenated coding scheme of polar codes. As shown in Figure  \ref{Fig7_CRC_Polar_ENC}, CRC-polar encoder consists of the CRC encoder and the polar encoder. $k$-bit source block  $c_1^k$ is input into the CRC encoder and generates the coded block $u_1^K$, where $K=k+m$ is the block length and $m$ is the CRC bits. These coded bits are mapped into the information bit set $\mathcal{A}$, i.e., $u_1^K=u_{\mathcal{A}}$. Then the information bits $u_\mathcal{A}$ and frozen bits $u_\mathcal{A}^c$ are fed into the polar encoder and generates the final concatenated codeword $x_1^N$. So the entire code rate is defined as $R=k/N$.

Thus the encoding process of CRC-polar codes can be written as
\begin{equation}
\left\{
\begin{aligned}
&u_1^N\mathbf{G}_N=x_1^N,\\
&u_\mathcal{A}=c_1^k\mathbf{G}_{CRC},u_\mathcal{A}^c=\mathbf{0}.
\end{aligned}
\right.,
\end{equation}
where $\mathbf{G}_{CRC}$ is the generator matrix of CRC codes.
\begin{figure}[htbp]
  \centering{\includegraphics[scale=0.7]{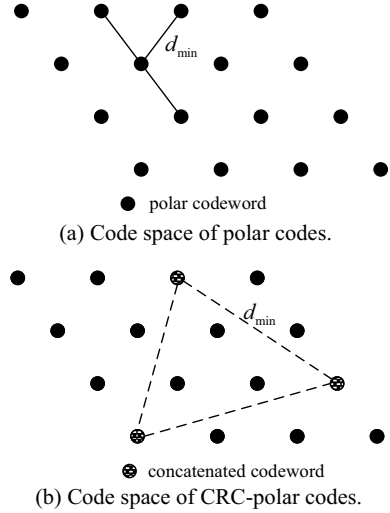}}
  \caption{Minimum Hamming distance comparison between polar code and CRC-Polar code.}\label{Fig8_Hamming_distance}
\end{figure}

The key reason that CRC-polar codes obtain the performance gain is CRC codes can significantly improve the minimum Hamming distance and the weight distribution of the polar codes. This principle can be intuitively depicted in Figure \ref{Fig8_Hamming_distance}. When single polar code is encoded in Figure 8(a), the Hamming distance of neighbour codewords, i.e., $d_{min}$ is small. On the contrary, when the concatenated coding is utilized, as shown in Figure 8(b), the minimum Hamming distance can be enlarged since the CRC check can exclude many neighbour candidates.

For the CRC-polar codes, the BLER probability can be upper bounded by the union bound \cite{Book_Lin}, that is,
\begin{equation}
\begin{aligned}
P_e & \leq \sum_{w=d_{min}}^{N}A_wQ\left(\sqrt{\frac{2wRE_b}{N_0}}\right)\\
      & \approx A_{d_{min}} Q\left(\sqrt{2d_{min}\frac{RE_b}{N_0}}\right),
\end{aligned}
\end{equation}
where $A_w$ is the weight enumerator and $\frac{E_b}{N_0}$ is the bit SNR.
We can find that the weight distribution $\{w, A_w\}$, especially the minimum weight $d_{min}$ and the enumerator $A_{d_{min}}$ will determine the performance of CRC-polar codes.

Given the code length $N=128,256$ and the code rate $R=k/N=1/2$, we compare the weight distribution of polar and CRC-polar codes as shown in Table \ref{WD_comp1} and Table \ref{WD_comp2}.
\begin{table*}[tp]
\centering
\caption{Weight distribution of Polar and CRC-Polar codes for $N=128$.} \label{WD_comp1}
\begin{tabular}{|c|c|c|c|c|c|}
\hline \multirow{2}{*}{Code type} &   \multirow{2}{*}{$K$}  & \multicolumn{4}{c|}{$A_w$}\\
\cline{3-6}                                      &                                       & $w=8$ & $w=12$ & $w=16$ & $w=20$ \\
\hline  Polar                                    &    64                                &    432   &	2304	    &  232440 & 1044823 \\
\hline  CRC6+Polar                        &  64+6                              &     4	    &     327   &   1301   & -\\
\hline  CRC6-opt+Polar                  &  64+6                              &     0	    &    300	 &     972	 & - \\
\hline
\end{tabular}
\end{table*}

\begin{table*}[tp]
\centering
\caption{Weight distribution of Polar and CRC-Polar codes for $N=256$.} \label{WD_comp2}
\begin{tabular}{|c|c|c|c|c|c|}
\hline \multirow{2}{*}{Code type} &   \multirow{2}{*}{$K$}  & \multicolumn{4}{c|}{$A_w$}\\
\cline{3-6}                                      &                                       & $w=8$ & $w=16$ & $w=20$ & $w=24$ \\
\hline  Polar                                    &    128                              &    96  	& 131824	&  548864  & 119215 \\
\hline  CRC9+Polar                        &  128+9                            &     0	    &  539	    &   2357     & -\\
\hline  CRC9-opt+Polar                  &  128+9                            &     0	    &  507	    &   1946	  & - \\
\hline  CRC10+Polar                      &  128+10                           &     0	    &  552      &   -          & -\\
\hline  CRC10-opt+Polar                &  128+10                           &     0	    &  215	    &   -      	 & - \\
\hline
\end{tabular}
\end{table*}
In these two tables, the optimal CRC codes, such as CRC6-opt etc., are selected from \cite{CRC_opt_Zhang}. In Table \ref{WD_comp1}, the generator polynomials of CRC6 and CRC6-opt are $g(x)=$ 0x43 and 0x73. Similarly, in Table \ref{WD_comp2}, the generator polynomials of CRC9, CRC9-opt, CRC10 and CRC10-opt are $g(x)=$ 0x2CF, 0x269, 0x633, and 0x75F. Here, we use the hexadecimal to present the generator polynomial.

We see that (128,64) polar code has the minimum Hamming weight $d_{min}=8$ and the weight enumerator $A_{d_{min}}=432$. When the CRC6-opt is applied, the $d_{min}$ of CRC-polar code is increased to 12 and the weight enumerator $A_{12}=300$ is also smaller than that of polar code, that is, $A_{12}=2304$. Similar phenomena can also be observed in Table \ref{WD_comp2}. When the CRC codes are used, the minimum Hamming distance will increase from 8 to 16. In a word, CRC concatenated coding is an efficient and powerful method to improve the performance of polar codes.

\subsection{Rate Compatible Polar Codes}
According to the availability of prior information, the rate-compatible polar codes can be divided into two modes. In the puncturing mode, some code bits are deleted in the encoder whereby these bits are unknown by the decoder and treated as the ones transmitting over zero-capacity channels. In the shortening mode, the values of deleted code bits are predetermined and known to both the encoder and decoder. Thus, the associated channels can be regarded as one-capacity channels.

Theoretically, the optimal puncturing table of rate compatible polar codes can be optimized by a brute-force search of the distance spectrum (for ML decoding) or BLER bounds (for SC decoding). However, the exhausted search for all the puncturing/shortening patterns is difficult to be realized.

Quasi-uniform puncturing (QUP) \cite{QUP_Niu} and reversal quasi uniform shortening (RQUS) \cite{Shorten_WangLiu}\cite{RateMatch_Niu} are two simple and high efficient schemes to achieve the optimal tradeoff between the error performance and implementation complexity of RCP codes.

Since the code length of classical polar codes is limited to a power of $2$, in order to generate an arbitrary $M$-bit length RCP code, we should start from an extended code of length $N=2^n$, with $n$ determined by $n=\lceil{\log M}\rceil$. Then an $N$-bit codeword is shrunk to $M$ bits by appropriately deleting $Q=(N-M)$ bits from it.

Define the puncturing/shortening table $\mathscr{T}_N=(t_1,t_2,\cdots, t_N)$ where $t_i\in\{0,1\}$ with $t_i=0$ meaning that the corresponding bit $x_i$ is deleted. For QUP scheme, the $N$-length table $\mathscr{T}_N$ is first initialized as all ones, and then set the \emph{first} $Q$ bits as zeros.
After the bit-reversal permutation on the table, the indices of zero elements in $\mathscr{T}_N$ should be punctured. Similarly, for RQUS, table $\mathscr{T}_N$ is first initialized as all ones and set the \emph{last} $Q$ bits as zeros. By the bit-reversal permutation, the indices of zero elements in $\mathscr{T}_N$ should be shortened.

\begin{figure*}[htbp]
  \centering{\includegraphics[scale=1]{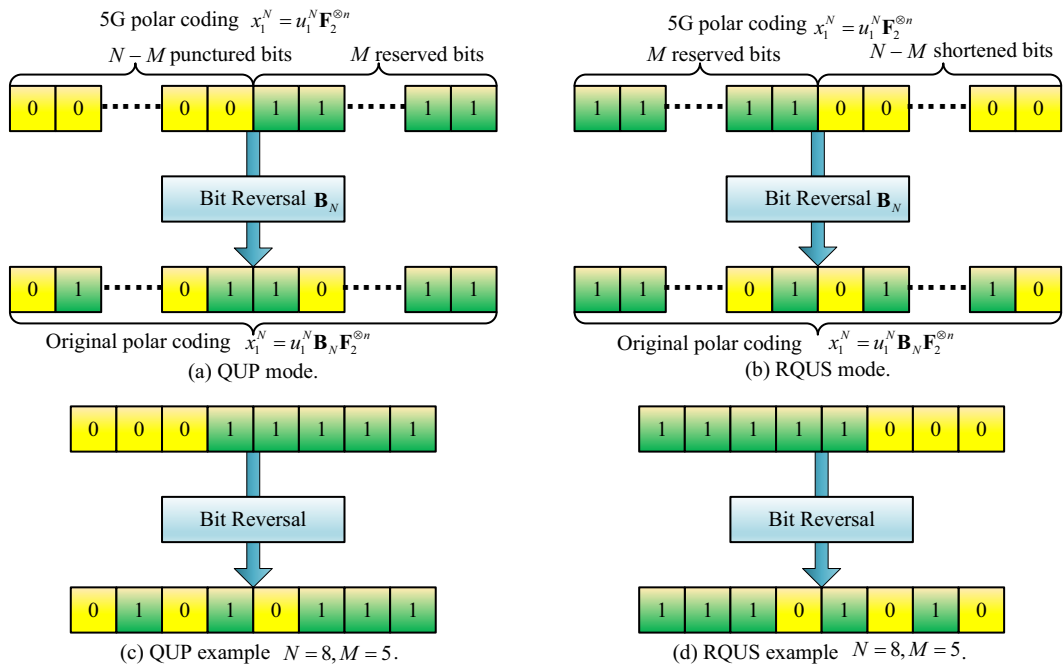}}
  \caption{QUP and RQUS scheme for rate compatible polar codes.}\label{Fig9_QUP_RQUS}
\end{figure*}

Figures 9(a) and 9(b) show the process of QUP and RQUS. Note that, since the natural order coding \eqref{natural_encoder} is used in 5G standard, the bit-reversal permutation is not necessary. Hence, QUP/RQUS scheme is equivalent to puncture (shorten) the first (last) $Q$ bits in the original codeword as shown in these two subgraphs. Since QUP is suitable for low code rate and RQUS is suitable for high code rate, 5G standard \cite{5GNR_38212} specifies that QUP is applied in the case of $R\leq7/16$ and RQUS is utilized in the case of $R>7/16$.

Figures 9(c) and 9(d) give the examples of QUP and RQUS respectively. Given $N=8, M=5, Q=3$, after bit-reversal permutation, the puncturing table is $\mathscr{T}_8=\left( 0,1,0,1,0,1,1,1 \right)$, that means the code bits $x_1$, $x_3$, and $x_5$ should be punctured. Similarly, the shortening table is $\mathscr{T}_8=\left( 1,1,1,0,1,0,1,0 \right)$, that means the code bits $x_4$, $x_6$, and $x_8$ should be shortened.
\section{Improved Polar Decoding Algorithms}
\label{section_V}
In this section, we briefly describe the basic principle of improved decoding algorithms of polar codes, such as SCL, SCS, SCH and SCP decoding. Finally, we demonstrate the superior performance of CA-SCL decoding by comparing the 5G polar codes, turbo and LDPC codes.

\subsection{Code Tree and SC Decoding}
\begin{figure}[htbp]
  \centering{\includegraphics[scale=0.7]{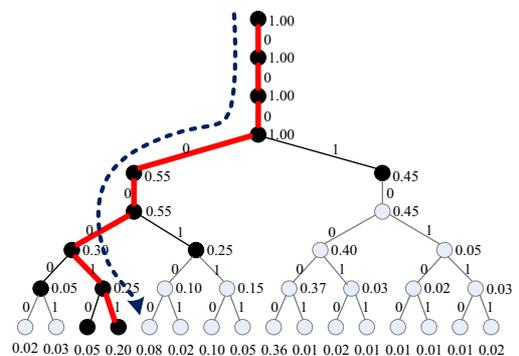}}
  \caption{SC decoding example on the code tree.}\label{Fig10_SC_code_tree}
\end{figure}

In \cite{Survey_Niu}, we introduced the compact-stage code tree to uniformly describe SC decoding and its improved algorithms, such as SCL/SCS/SCH/SCP decoding. Figure \ref{Fig10_SC_code_tree} gives an example of SC decoding over a compact-stage code tree, which is corresponding to the trellis of Figure \ref{Fig4_SC_decoder_structure} and consists of eight levels by compacting the stages.

In this code tree, except the leaf nodes and the frozen nodes, each node has two descendants and the corresponding branches are labeled with 0 and 1, respectively. The number written next to each node denotes the path metric from the root to that node, e.g. LLR or a posteriori probability (APP). A decoding path includes a series of branches from the root to one leaf node. In Figure \ref{Fig10_SC_code_tree}, the black circles represent the nodes that are visited and the gray ones are those that are not visited in the search process.

Since at a certain level associated with an information bit only one of the two branches with the larger/better path metric is selected for further extension, the SC decoding can be regarded as a greedy search over the compact-stage code tree. Hence, if a decision error occurs, this error will propagate in the extended path. As shown in Figure \ref{Fig10_SC_code_tree}, the SC decoding path ``00000011'' (marked by the red bold branches) is not optimal due to the level by level decision strategy.

\subsection{Successive Cancellation List Decoding}
In order to mitigate the error propagation effect and improve the performance of SC decoding, Chen and Niu  \cite{SCL_Conf}\cite{SCL_JNL}, Tal and Vardy \cite{Tal_Vardy_ISIT}\cite{Tal_Vardy_TIT} independently proposed the SCL decoding. In fact, Dumer \cite{List_Dumer} proposed the similar idea in order to improve the performance of RM codes.

Unlike SC decoder only reserves one path at each level, SCL decoder can extend $2L$ paths at each level and select $L$ most reliable paths as the survivors. In the end, the path with the largest metric is selected from the survivor list as the final decision. The time complexity of SCL decoding is $O(LN\log_2N)$ and the space complexity is $O(LN)$.

\begin{figure}[htbp]
  \centering{\includegraphics[scale=0.7]{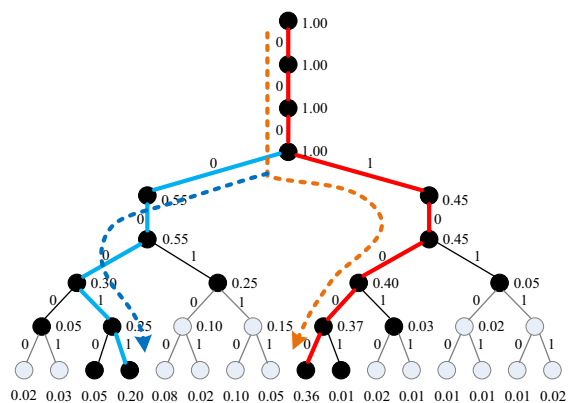}}
  \caption{SCL decoding example on the code trees.}\label{Fig11_SCL_code_tree}
\end{figure}

Figure \ref{Fig11_SCL_code_tree} depicts an example of SCL decoding with list size $L=2$. In each level, four paths are extended and the corresponding path metrics are calculated. After path sorting, two survivor paths (marked by blue and red bold edges) are selected and stored. Finally, two survivors are stored in the list. One path ``00000011'' with path metric 0.20 and the other path ``00010000'' with metric 0.36. Due to $0.36>0.20$, the more reliable path ``00010000'' is found whereby the error propagation in SC can be efficiently weaken.

\subsection{Successive Cancellation Stack Decoding}
SCL decoding substantially improves the performance of finite length polar codes yet the computational complexity is also increased. In order to reduce the decoding complexity, Niu \emph{et al.} \cite{SCS_Niu} proposed the SCS algorithm by using the depth-first search over the code tree. The SCS decoder goes along the code tree and uses a stack to sort and store the candidate paths. First, the top path is extended. Then, the succeeded paths are sorted and inserted into the stack. Until the top path with the largest metric reaches a leaf node, the decoding process stops and this path is output as the optimal decision. The main difference between SCL and SCS is that the candidate path of the former has the same length whereas the path of the latter has distinct length. Since SCS decoder only extends and calculates the necessary path nodes, its time complexity is far bellow that of SCL decoder even close to SC decoder. On the other hand, in order to achieve the same performance as SCL decoder, SCS needs a large stack depth $D$ and the space overhead is $O(DN)$. In the worst case, the space complexity is up to $O(LN^2)$ with $D=LN$.

\begin{figure}[htbp]
  \centering{\includegraphics[width=1.0\columnwidth]{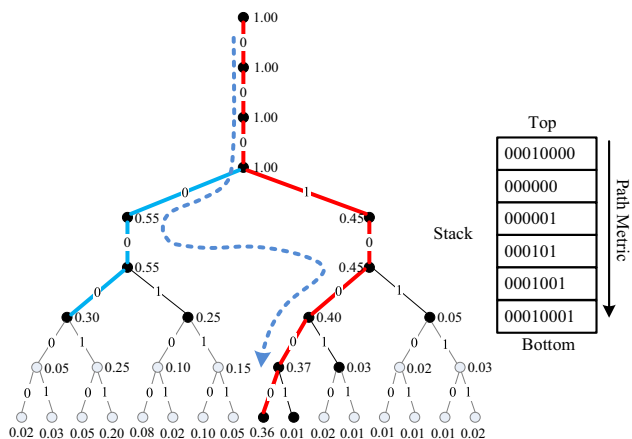}}
  \caption{SCS decoding example on the code tree.}\label{Fig12_SCS_code_tree}
\end{figure}
Figure \ref{Fig12_SCS_code_tree} gives a simple example of SCS decoding. In the stack, the top path (marked by red bold edges) has the largest metric 0.36 and the second candidate (marked by blue bold edges) has the second largest metric 0.3. Obviously, the top path ``00010000'' is output as the optimal decision, which is the same as the decoding result of SCL. However, the length of the top path is 8 and that of the second path is 6. Therefore, the number of the visiting nodes in SCS is small than that in SCL. So the time complexity of SCS is lower than that of SCL.

In order to make a tradeoff between the time complexity and the space overhead, Chen and Niu \emph{et al.} devised the successive cancellation hybrid (SCH) decoding in \cite{SCH_Chen} by combining SCL and SCS. This algorithm can achieve the same performance as SCL and SCS with a lower time and space complexity.

\subsection{Successive Cancellation Priority Decoding}
Guan and Niu \emph{et al.} \cite{SCP_Guan} proposed another algorithm, named successive cancellation priority (SCP) decoding, to reduce the time complexity of SCL. The SCP decoder performs priority-first decoding by interacting the priority queue with the trellis iteratively. On one hand, the priority information is stored in the priority queue so as to guide the extension of the candidate path. On the other hand, the trellis calculates and stores the intermediate results. Due to reducing most of the unnecessary path extensions by using the priority queue, the time complexity of the SCP decoder is much lower than the standard SCL decoder.
\begin{figure}[htbp]
  \centering{\includegraphics[scale=0.7]{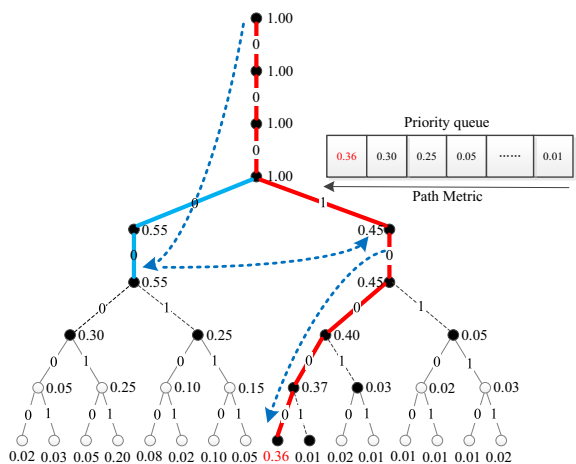}}
  \caption{SCP decoding example on the code tree.}\label{Fig13_SCP_code_tree}
\end{figure}

An example of SCP decoding is illustrated in Figure \ref{Fig13_SCP_code_tree}. In the priority queue, the survivor path (marked by red bold edges) has the largest metric 0.36 and ranked at the queue head. According to the priority information, the trellis calculates and extends the candidate paths. Hence, the survivor path ``00010000" is also output as the optimal decision. However, since only the candidate paths in the queue are extended in the trellis, the time complexity of SCP is lower than that of standard SCL.

We summarize the characteristics of the four improved SC algorithms in Table \ref{SCdecoding_summary}. All the four algorithms can approach the performance of ML decoding. SCL is the baseline to improve the performance of SC decoding with a computational complexity $O(LN\log_2N)$. SCS is the counterpart to achieve the same performance with a low computational complexity $O(N\log_2N)$ in high SNR yet high space complexity $O(LN^2)$. As a hybrid decoding, SCH can keep a good balance between the computational and space complexity. Finally, SCP presents another technique to achieve a good tradeoff.
\begin{table*}[htbp]
\centering
\caption{Summary of improved SC algorithms.} \label{SCdecoding_summary}
\begin{tabular}{|c|c|c|c|c|}
\hline  Algorithm    &  Searching strategy    &  Error Performance     &    Computational complexity           &  Space complexity    \\
\hline     SCL         &     width-first             &    approaching ML      &       $O(LN\log_2N)$                      &        $O(LN)$           \\
\hline     SCS         &     depth-first             &    approaching ML      &      $O(N\log_2N)$ in high SNR         &        $O(LN^2)$           \\
\hline     SCH         &     width/depth          &    approaching ML      &      $O(N\log_2N)$ in high SNR          & $O(LN)\sim O(LN^2)$ \\
\hline     SCP         &   trellis/priority queue & approaching ML      &       $O(N\log_2N)$ in high SNR          &        $O(LN\log_2 N)$    \\
\hline
\end{tabular}
\end{table*}

All these improved SC decoding algorithms can be applied in CRC-polar codes. In this case, the SCL/SCS/SCH/SCP decoder outputs the candidate paths into a CRC detector, and the check results are utilized to detect the correct codeword. Using these CRC-aided decoding schemes, the performance of CRC-polar codes can substantially outperforms that of turbo/LDPC codes.

\begin{figure}[htbp]
  \centering{\includegraphics[width=1.0\columnwidth]{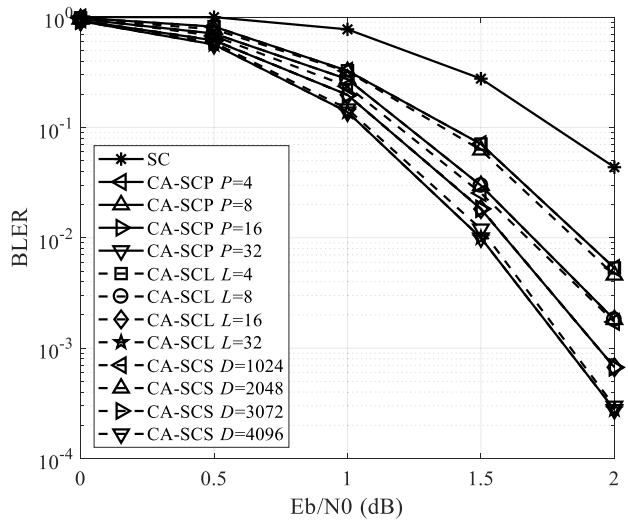}}
  \caption{Performance comparison of SC, CA-SCL, CA-SCS and CA-SCP decoding for (1024,512) CRC-polar code in AWGN channel.}\label{Fig14_BLER_comparison}
\end{figure}

We compare the BLER performance of SC, CA-SCL, CA-SCS and CA-SCP decoding for (1024,512) CRC-polar code under the AWGN channel in Figure \ref{Fig14_BLER_comparison}. The polar code is constructed by GA algorithm and CRC8 codes is used with the generator polynomial $g(x)=$ 0x9F. We can see that the SC decoding has the worst performance. When the search width $P$ in CA-SCP is configured as the same as the list size $L$ in CA-SCL, these two decoders achieve the same performance. Furthermore, when the CA-SCS decoder is set to a large depth stack (e.g. $D=1024$), it can also approach the same performance as CA-SCL or CA-SCP.

\subsection{Performance Evaluation of 5G Polar Codes}
\begin{figure*}[htbp]
  \centering
  \includegraphics[scale=0.51]{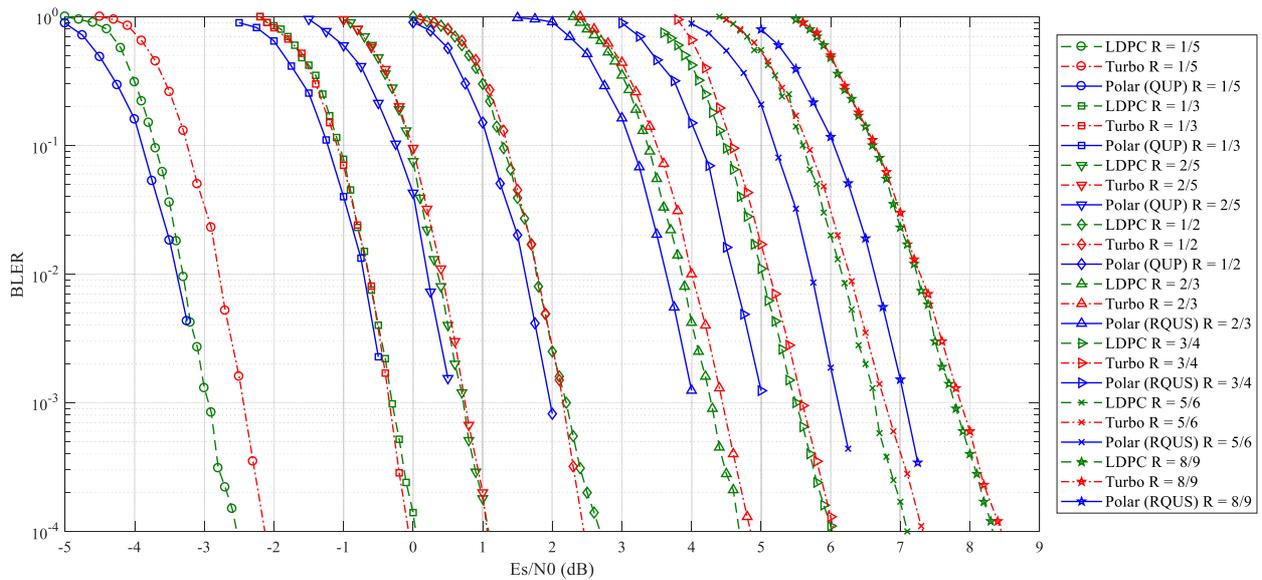}
  \caption{BLER performance comparison of RCP codes (punctured by QUP and RQUS algorithms), LTE turbo codes and LDPC codes with information length $K=400$ and various code rates.}
  \label{Fig15_performance_comp_400}
\end{figure*}

In 5G standard, CRC concatenated rate-compatible polar codes (constructed by QUP or RQUS algorithms) are the basic schemes of polar codes. Next we compare the performance of 5G polar codes, turbo and LDPC codes under the AWGN channel. The BLER curves vs $E_s/N_0$ with the information length $K=400$ and code rate $R=1/5\sim8/9$ are shown in Figure \ref{Fig15_performance_comp_400}. 5G polar codes are constructed from the parent code with the code length $N=1024$ by QUP or RQUS schemes. The CA-SCL decoding algorithm is employed with the list size $L=32$ and a CRC8 code with the generator polynomial $g(x)=$ 0x9F is used. An eight-state turbo code in 3GPP LTE standard \cite{LTE_36212} is used as a reference. The LDPC codes proposed by Qualcomm corporation in the 5G standard proposal of channel coding \cite{Qualcomm_Proposal} are applied. The logarithmic maximum a posterior (Log-MAP) algorithm is applied in turbo decoding and the maximum number of iterations is ${{I}_{\max }}=8$. And the belief propagation (BP) algorithm is applied in LDPC decoding and the maximum number of iterations is ${{I}_{\max }}=50$.

In most cases, polar codes can achieve additional coding gains relative to turbo or LDPC codes. In the region of low code rate $R=1/5\sim1/2$, as shown in Figure \ref{Fig15_performance_comp_400}, the polar codes punctured by QUP can achieve the same or slightly better performance than those turbo or LDPC codes. Especially for the code rate $R=1/5$, polar codes can outperform the turbo codes by $0.5\sim1$dB. Compared with turbo or LDPC codes, polar codes are a kind of powerful channel codes at the short to medium code length and the performance gain can be further improved by increasing the list size.

In the region of high code rate $R=2/3\sim8/9$, compared to turbo codes, a maximum $1\sim1.2$ dB additional gains can be obtained at the code rate $R=8/9$. On the other hand, compared to LDPC codes, a maximum $0.8$dB performance gain can be attained at the code rate $8/9$. In contrast to the case of low code rate, the RQUS scheme can generate better polar codes in this case. In addition, both turbo and LDPC codes show an error floor phenomenon with the increasing SNR. On the contrary, CRC-polar codes have no such effect.

\begin{remark}
In 5G standard \cite{5GNR_38212}, either the downlink (downlink control indicator (DCI) or broadcast channel (BCH)) or uplink (uplink control indicator (UCI)) message bits are encoded by the CRC-polar codes and QUP/RQUS rate matching schemes. Accordingly, these control singling or channels are dictated the CA-SCL decoding \cite{CASCL_Niu} as the channel decoder algorithm to provide high reliability. As commented by Matlab 5G toolbox \cite{5G_toolbox}, ``it is well known that CA-SCL decoding can outperform turbo or LDPC codes and this was one of the major factors in the adoption of polar codes by 3GPP".
\end{remark}

\section{Polar Coding in the Future}
\label{section_VI}
In the future 6G system, ultra-high reliable transmission, high spectrum efficiency and large system capacity are the primary requirements of wireless communications. In this section, we mainly discuss the polar coding to fulfill these requirements. First, we briefly introduce the limited performance of short polar code to satisfy the reliable transmission. Second, the code construction methods are investigated to improve the performance of polar codes under the fading channels. Third, polar coded modulation and polar coded HARQ are discussed in order to increase the spectrum efficiency. In the end, we design the framework of polar processing to jointly optimize the wireless transmission systems.
\subsection{Optimal Short Polar Codes}
Low-latency ($100\mu s$) and ultra-high reliability (BLER$<10^{-6}\sim10^{-7}$) are the key performance metrics of 6G wireless transmission \cite{6G_whitepaper}. These requirements become the serious challenges for the short channel codes.

In 2019, Ar{\i}kan \cite{PAC} proposed the polarization-adjusted convolutional (PAC) codes with the Reed-Muller design rule to improve the ML performance of short polar codes. PAC codes have excellence performance and approach the normal approximation (NA) of the finite blocklength capacity in \cite{PPV_bound}. However, the code rate is not flexibly configured and the complexity of sequential decoding is very high when the code rate approaches the capacity.

Piao and Niu \emph{et al.} \cite{CAHD} found that the simple CRC-polar codes by the optimization of encoding and decoding can also approach the normal approximation. First, they used the sphere constraint based enumeration methods \cite{SCBEM} to analyze the minimum weight distribution of CRC-polar codes. In fact, in the case of short code length, the generator polynomial of CRC codes is very significant to affect the performance of CRC-polar codes. Table \ref{CRC_polar_poly} gives the optimal generator polynomials of CRC codes for short CRC-polar codes with code length $N=128$. Given the code rate $R=1/3,1/2,2/3$ and the CRC bit length $m=20,24,16$, the optimal polynomials $g(x)$ (indicated by hexadecimal) listed in this table are obtained after the brute-force searching by calculating the minimum weight distribution, where $d_{min}$ is the minimum Hamming distance and $A_{d_{min}}$ is the corresponding enumerator.

\begin{table}[htbp]
\centering
\caption{Optimal generator polynomials of CRC-Polar codes with code length $N=128$.} \label{CRC_polar_poly}
\begin{tabular}{|c|c|c|c|c|c|}
\hline            $N$                     &  $R$   &  $m$     &    $g(x)$          &      $d_{min}$   &     $A_{d_{min}}$\\
\hline  \multirow{3}{*}{128}   &   1/3    &    20      &   0x1005D1     &           24           &        171                \\
\cline{2-6}                              &   1/2    &    24	      &   0x10001E5	&           16           &         66                 \\
\cline{2-6}                              &   2/3    &    16	      &   0x117B7	    &           10	         &        167                \\
\hline
\end{tabular}
\end{table}

\begin{figure}[htbp]
  \centering
  \includegraphics[scale=0.7]{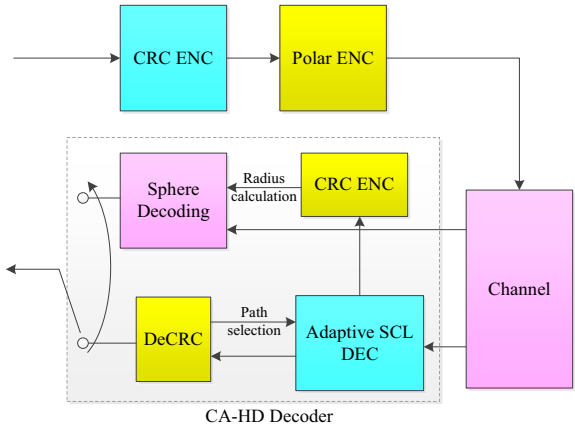}
  \caption{CA-HD decoding for CRC-polar codes.}
  \label{Fig16_CAHD_Scheme}
\end{figure}

Then, they designed a new CRC-aided hybrid decoding (CA-HD) for the CRC-polar codes by combining the CA-SCL and CA-SD algorithm. Figure \ref{Fig16_CAHD_Scheme} shows the diagram of CA-HD decoding, which includes two decoding modes: Adaptive CA-SCL mode and CA-SD mode.

As shown in Figure \ref{Fig16_CAHD_Scheme}, the adaptive CA-SCL decoding is first started to estimate the transmitted codeword. If the list size is smaller than the maximum size $L_{max}$ and the decision results can pass the CRC check, then the decoding is terminated. Otherwise, when the list size reaches the maximum size and the decoding is failed, CA-SD decoding is activated. In order to reduce the searching radius of sphere decoding, the CRC bits are re-encoded and input into the SD decoder. In the end, the result of SD decoding is output as the final decision. CA-SCL decoding has low complexity but the performance is inferior to the ML decoding. On the contrary, CA-SD decoding is identical to the ML decoding whereas the complexity is very high. Therefore, by elaborately integrating CA-SCL and CA-SD, the CA-HD decoding can achieve the optimal tradeoff between the performance and complexity.

\begin{figure}[htbp]
  \centering
  \includegraphics[width=1.0\columnwidth]{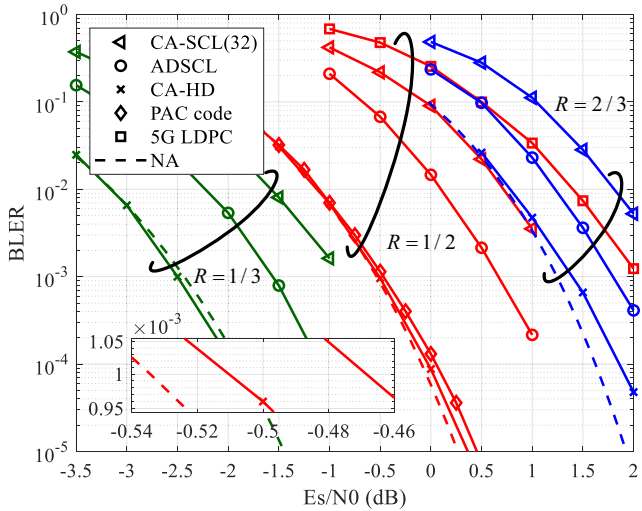}
  \caption{The performance comparison of CRC-polar codes with various decoding algorithms and PAC codes at the code length $N=128$.}
  \label{Fig17_CAHD_BLER}
\end{figure}

We investigate the performance of CRC-polar codes with the code length $N=128$ and code rate $R=1/3$, $1/2$, and $2/3$ (configured parameters are selected from Table \ref{CRC_polar_poly}) under various algorithms, such as ADSCL (the maximum list size $L_{max}=1024$), CA-SCL (the fixed list size $L=32$) and CA-HD ($L_{max}=1024$). Figure \ref{Fig17_CAHD_BLER} depicts the BLER performance of CRC-polar codes. Furthermore, the BLER performance of (128,64) PAC code, LDPC code and the NA bound are also illustrated in Figure \ref{Fig17_CAHD_BLER}. Here, the performance of LDPC code is evaluated based on the construction of 5G NR LDPC code with the base graph 2 (BG2). and the belief propagation (BP) decoding algorithm by 50 iterations.

We observe that the CRC-polar code under CA-SCL and ADSCL can achieve better performance than LDPC code. Furthermore,
the performance under CA-HD decoding significantly outperform that under CA-SCL and ADSCL and approach the NA bound. Especially, for the (128,64) code, the CRC-polar code is superior to PAC code and LDPC code and very close to the NA bound with a small gap of $0.025$ dB. We can conclude from these results that CRC-polar codes with the optimized generator polynomial and CA-HD decoding can almost approach the finite blocklength capacity. Therefore these optimal CRC-polar codes can be regarded as the important candidate to provide the low-latency and ultra-high reliability transmission in the 6G wireless system.

On the other hand, for the medium blocklength, the performance of CRC-polar codes is evaluated and compared with turbo/LDPC codes in \cite{Survey_Niu}. For an example, given the code length $N=1024$ and the code rate $R=\frac{1}{2}$, the CRC-polar code can achieve $0.5\sim1$ dB performance gain over the turbo/LDPC code at the BLER of $10^{-4}$. Generally, for the moderate to long blocklength, polar codes under the SCL decoding with a large list size will reach a similar or better performance than the turbo/LDPC codes. In addition, CRC-polar codes show no sign of error floors in the high SNR regime, which is a significant advantage over the turbo/LDPC codes.
\subsection{Construction in Fading Channels}
The construction of polar codes in the fast or block fading channels is an important direction for the practical application and has been attracted wide attentions. For the fast Rayleigh fading channel, Trifonov \cite{Fading_Trifonov} first presented an iterative algorithm to calculate and track the diversity order and noise variance of the polarized channels. Lately, Zhou and Niu \emph{et al.} \cite{Fading_Zhou} designed two algorithms to find a capacity-equivalent BI-AWGN channel of the Rayleigh channel and constructed the polar codes by using the GA algorithm. Recently, Niu and Li \cite{FastFading_Niu} established a systematic framework in term of the polar spectrum to analyze and construct polar codes in various fast fading channels, such as Rician, Rayleigh and Nakagami channels, which explicitly reveals the relationship between the diversity order and the codeword weight.

On the other hand, for the block fading channel, Bravo-Santos \cite{Fading_Bravo-Santos} proposed a recursive calculation of Bhattacharyya parameter yet with a time-consumption Monte-Carlo computation. Subsequently, Si \emph{et al.} \cite{Fading_Si} designed a two-stage polar coding over channel uses and fading blocks. However the construction still depends on the recursive calculation of Bhattacharyya parameter. Niu and Li \cite{BlockFading_Niu} systematically analyzed the error performance of polar codes by introducing the new concept, named split polar spectrum. The upper bound on the error probability of polarized channel is derived to explicitly reveal the relationship between the diversity order $L$ and the block-wise weight distribution of the codeword.

In a word, the polar codes constructed based on polar spectrum \cite{FastFading_Niu}\cite{BlockFading_Niu} can achieve similar or better error performance than those constructed by conventional construction. Due to the advantages of low complexity and high performance, these methods are suitable for the construction of polar codes in wireless communications.

\subsection{Polar Coded Modulation and HARQ}
In order to fulfill the high spectrum efficiency requirement of 6G wireless communications, extending the concept of channel polarization and designing the polar coded modulation (PCM) and polar coded hybrid automatic repeat request (PC-HARQ) system become the key technologies in wireless data transmission.

\begin{figure}[htbp]
  \centering
  \includegraphics[scale=0.7]{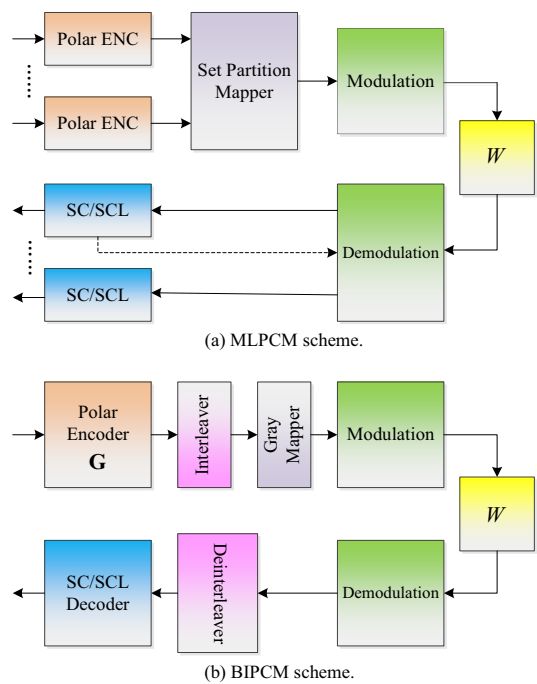}
  \caption{System architecture of PCM.}
  \label{Fig18_System_Architecture_PCM}
\end{figure}

Seidl \emph{et al.} \cite{PCM_Seidl} first established the framework of polar-coded modulation, which includes two-stage channel polarization transform, such as coding polarization and modulation polarization. The PCM framework can be designed based on two coded-modulation schemes, i.e., the bit-interleaved polar-coded modulation (BIPCM) and multi-level polar coded modulation (MLPCM). Figure \ref{Fig18_System_Architecture_PCM} shows the system architectures of two schemes. For the MLPCM, multiple polar encoders are utilized and a tuple of coded bits is directly mapped to a modulated symbol. On the other hand, for the BIPCM, a single polar encoder is used and the coded bit sequence is interleaved and mapped to the modulated symbol. Generally, the bit-to-symbol mapping is very important for the PCM design in order to enhance the polarization effect. Gray mapping and set partition mapping are suitable for BIPCM
and MLPCM respectively. Furthermore, rate-matching (e.g. QUP scheme) and coding construction (e.g. GA algorithm) are also critical techniques in the PCM design.

For the MLPCM, Zhou and Niu \emph{et al.} \cite{Univ_MLPCM} extended the PW metric and presented a universal construction for MLPCM to facilitate the practical implementation. Then Khoshnevis \emph{et al.} \cite{Throughput_MLPCM} established the throughput maximization scheme by using the set-partition (SP) mapping and the rate matching algorithm. Furthermore, Dai \emph{et al.} \cite{Asyn_MLPCM} introduced the spatial coupled structure among multiple coded blocks and designed the asynchronous polar-coded modulation scheme to improve the transmission reliability of MLPCM.

On the other hand, for the BIPCM, Shin \emph{et al.}\cite{Mapping_Shin} found the mapping patterns for the pulse-amplitude modulation (PAM) with Gray labeling whereas the search complexity is very high. Then, using the constellation symmetry, Chen and Niu \emph{et al.} \cite{BIPCM_ChenNiu} proposed an efficient search algorithm to find the optimal mapping of BIPCM. In order to improve the performance of BIPCM, Tian \emph{et al.} \cite{JSCD_BIPCM} designed a joint successive cancellation decoding algorithm by combining the demapping and deinterleaving into SC decoder. Mahdavifar \emph{et al.} \cite{PolarCoding_BIPCM} considered the multi-channel model of BICM and constructed the compound polar code.

Hybrid automatic repeat request (HARQ) is another key technology to enhance the link reliability and throughput in practical wireless systems. Chen and Niu first proposed two types of the polar-coded HARQ schemes, such as incremental redundancy (IR) PC-HARQ \cite{IRHARQ} and chase combining (CC) PC-HARQ \cite{CCHARQ}. By using the QUP scheme, the original IR-PC-HARQ scheme \cite{IRHARQ} can achieve the same or higher throughput than turbo/LDPC code HARQ but its latency is slightly high. On the other hand, the CC-PC-HARQ \cite{CCHARQ} is easy to implement while the system performance is limited due to small coding gain.

Lately, Li \emph{et al.} \cite{IFHARQ} proposed the incremental freezing (IF) scheme so as to achieve coding gain at retransmissions. Further, an adaptive IR scheme based on the polarizing matrix extension (PME) was proposed in \cite{MEHARQ}. By using QUP method, the PME based PC-HARQ scheme constructs a longer length polar code with multiple transmissions. Compared with the IF scheme, the PME scheme can obtain additional coding gain with enhanced polarization effect.

In summary, polar coded modulation and HARQ will become the key supports of the 6G wireless transmission. Considering the practical application, the construction method, rate-matching and mapping pattern are needed to be further explored in the future.

\subsection{Polar Processing}
Polarization effects exist in almost every unit of the communication system rather than only in the coding module. Theoretically, when the code length goes to infinite, polar coding can achieve the corresponding limitation of various communication scenario, such as loss/lossy source coding, multiple access, broadcasting, relay and distributed communication. As stated in \cite{Dissertation_Korada}, Korada pointed out that polarization is almost optimal for everything. From the viewpoint of practical application, Niu \emph{et al.} \cite{Polar_processing} proposed the framework of the polar coded transmission, name polar processing, as a new design methodology to fulfill the high spectrum efficiency of 6G wireless system.

\begin{figure}[htbp]
  \centering
  \includegraphics[width=1.0\columnwidth]{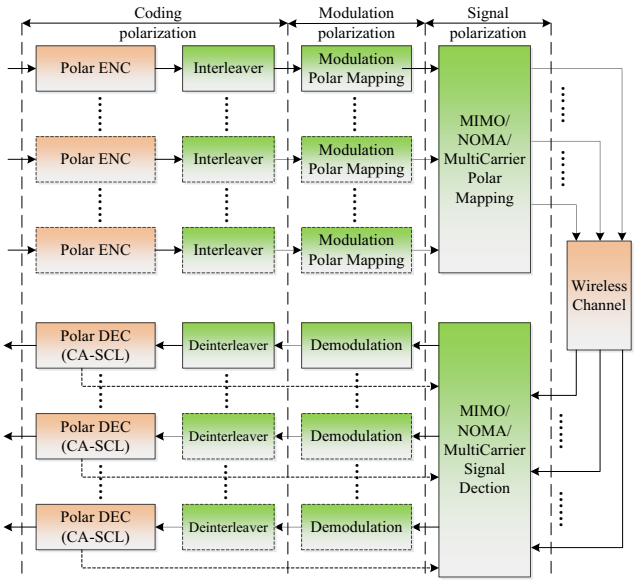}
  \caption{System architecture of polar processing.}
  \label{Fig19_Polar_Processing_Architecture}
\end{figure}

Figure \ref{Fig19_Polar_Processing_Architecture} illustrates the system architecture of polar processing. The architecture consists of three-stage polarization: coding polarization, modulation polarization and signal polarization. In the first stage, i.e., signal polarization, many signal processing techniques have a general polarization effect. For an example, in the multiple-input-multiple-output (MIMO) system, different antenna link has distinct channel reliability. Similarly, in non-orthogonal multiple access (NOMA) system, different user undergoes diverse channel condition. In multi-carrier system, similar phenomenon can be observed. Therefore, individual antenna/user/carrier can be regarded as the general polarized channel and the reliability distinction in each channel reveals the polarization in signal space. In the second stage, i.e., modulation polarization, one signal streaming is further decomposed into many modulated bit subchannels with different reliability. Finally, in the third stage, i.e., coding polarization, we use the one or multiple polar codes to match each modulated-bit polarized channels.

By using the three-stage polarization method, the polar processing transmitter glues the coding, modulation and signal processing into a joint polarization architecture so as to dramatically improve the system performance. Meanwhile, the polar processing forms a joint successive cancellation structure whereby CA-SCL decoding, soft demodulation and soft signal detection can be integrated into a low complexity compounded SCL receiver rather than the complex iterative calculation in turbo processing.

Under the framework of polar processing, Dai and Niu \emph{et al.} \cite{Polar_MIMO} designed the polar-coded MIMO (PC-MIMO) system by the bit/symbol/antennal polarization. Then they \cite{Polar_NOMA} proposed the polar-coded NOMA (PC-NOMA) system by the bit/symbol/user polarization. Lately, Li and Niu \emph{et al.} \cite{Polar_GFDM} investigated the polar-coded generalized frequency division multiplexing (GFDM) systems by the similar methods. Recently, Piao and Niu \emph{et al.} \cite{Polar_precoding} considered the polar-coded precoding system by designing a unitary finite-feedback transmit precoder.

\begin{figure}[htbp]
  \centering
  \includegraphics[width=1.0\columnwidth]{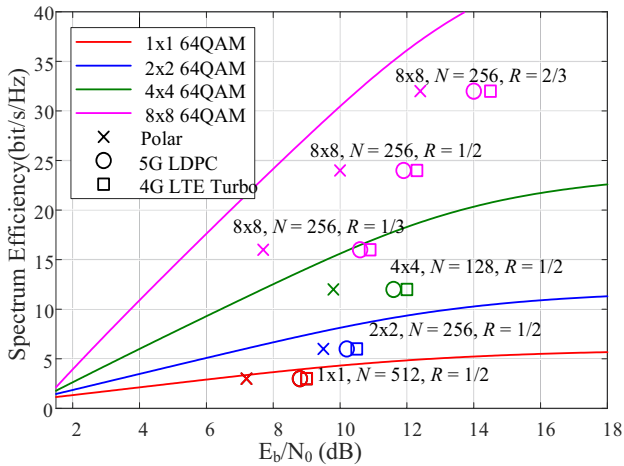}
  \caption{Performance comparison of PC-MIMO, TC-MIMO and LC-MIMO.}
  \label{Fig20_Polar_MIMO_BLER}
\end{figure}

Give different configurations of MIMO ($1\times1,2\times2,4\times4,8\times8$) and $64$ QAM modulation, in the case of BLER$=10^{-4}$, we evaluate the spectrum efficiency of three coded MIMO systems, that is, PC-MIMO, Turbo coded MIMO (TC-MIMO) and LDPC coded MIMO (LC-MIMO) in Figure \ref{Fig20_Polar_MIMO_BLER}. The polar codes are constructed by using the method in \cite{Polar_MIMO} and CA-SCL decoding is used. Turbo codes are referred to LTE standard \cite{LTE_36212} and the decoding is the Log-MAP algorithm. LDPC codes are utilized from 5G standard \cite{5GNR_38212} and the BP decoding is applied.

As shown in Figure \ref{Fig20_Polar_MIMO_BLER}, for all the MIMO configurations, PC-MIMO can achieve $1\sim2$ dB performance gain over TC-MIMO or LC-MIMO. Since the polar processing jointly polarized the entire communication system, PC-MIMO can significantly improve the spectrum efficiency. Therefore, it follows that PC-MIMO is a powerful candidate technology to fulfill the high efficiency transmission requirement of 6G system.

\begin{remark}
By now, polar codes have been adopted as the coding standard of control channels in 5G wireless system. Due to the constraints of the control channels, the concatenated coding, construction and rate-matching of polar codes are standardized while other advanced techniques are still open. The application in 5G is only a start point of polar codes in practical implementation. Look to the future, the polar coded transmission, or equivalently polar processing, will uncover a universal and powerful methodology to optimize the communication system. Due to the double advantages of performance and implementation, we believe that polar codes and polar processing will become more popular in the 6G wireless system, satellite communication system, microwave communication system, etc.
\end{remark}

\section{Tao and Polarization: the Traditional and the Modern}
\label{section_VII}
In this paper, we review the basic principle of polar codes, the application in 5G standard and the promising directions in the future. As a great breakthrough of theory, the invention of polar codes is an important milestone of information theory and channel coding to uncover the constructive method approaching the channel capacity. For the practical application, polar codes with concatenated coding and CA-SCL decoding fulfill the high reliability requirement of 5G wireless system at the short to medium code length. In the future, polar processing may guide a revolution of system design and open a new era of communication technology.

\begin{figure}[htbp]
  \centering
  \includegraphics[width=1.0\columnwidth]{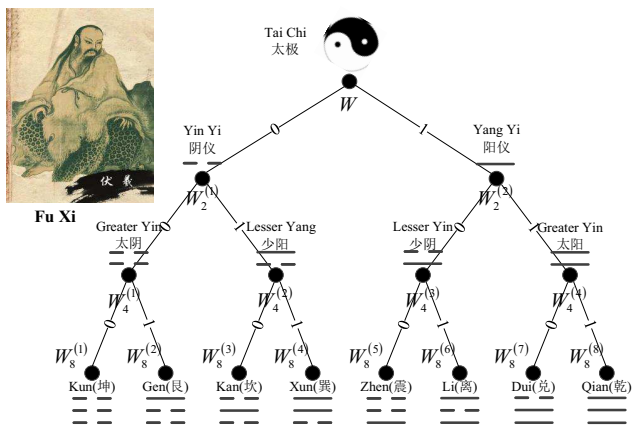}
  \caption{The eight trigrams and code tree.}
  \label{Fig21_Polar_code_Bagua}
\end{figure}

When we retrospect the Chinese traditional culture, we find the fantastic correspondence between the polarization and Tao. In the famous work of Taoism, I Ching \cite{IChing}, it explains the principle of change. That is to say, Tai Chi gives rise to Liang Yi, Liang Yi give rises to four phenomena (Si Xiang), Si Xiang give rises to Eight Trigrams, and the understanding of auspiciousness through Eight Trigrams. According to legend, the ancient Chinese figure of Fu Xi invented the eight diagrams. The modern binary number system, the basis for binary code, was invented by Gottfried Leibniz in 1689. He encountered the I Ching and noted that similar presentation between the binary numbers and the eight trigrams \cite{Leibniz}. Recently, we find the 1-1 mapping relationship between the eight trigrams and the code tree of polar codes as shown in Figure \ref{Fig21_Polar_code_Bagua}.

We observe that Tai Chi is mapped to the root node and Yin Yi and Yang Yi are associated to the polarized channels $W_2^{(1)}$ and $W_2^{(2)}$ respectively. Furthermore, four phenomena, such as, Greater Yin, Shao Yang, Lesser Yin and Greater Yin, are corresponding to four polarized channels $W_4^{(i)},i=1,2,3,4$. Finally, eight trigrams are 1-1 mapped to eight polarized channel $W_8^{(i)},i=1,2,\cdots,8$.  The eight trigrams indicate everything in the world and have distinct manifestation. On the other hand, the polarized channels have different reliabilities due to the channel polarization. It seems that I Ching provides an interesting interpretation of channel polarization. If we deeply explore the idea of Taoism, such ancient Chinese philosophical thought may guide a new insight into the design and optimization of the polar codes.

Passing through 3,000 years, the traditional Taoism and the modern coding contrast finely with each other. This is an amazing orchestration between the classic philosophy and modern technology!

\section*{Acknowledgement}
This work is supported in part the Key Program of National Natural Science Foundation of China (No. 92067202), in part by the National Natural Science Foundation of China (No. 62071058), and in part by the Major Key Project of PCL (PCL2021A15).

\theendnotes

\bibliographystyle{gbt7714-numerical}
\bibliography{myref}

\biographies
\begin{CCJNLbiography}{photo_Niu.eps}{Kai Niu}
received the B.S. and Ph.D. degrees from Beijing University of Posts and Telecommunications (BUPT), Beijing, China, in 1998 and 2003. He is currently a Professor with the School of Artificial Intelligence, BUPT. His research interests include channel coding theory and applications, semantic communication and broadband wireless communication.
\end{CCJNLbiography}

\begin{CCJNLbiography}{photo_Zhang.eps}{Ping Zhang}
is currently a professor of School of Information and Communication Engineering at Beijing University of Posts and Telecommunications, the director of State Key Laboratory of Networking and Switching Technology, a member of IMT-2020 (5G) Experts Panel, a member of Experts Panel for China’s 6G development. He served as Chief Scientist of National Basic Research Program (973 Program), an expert in Information Technology Division of National High-tech R\&D program (863 Program), and a member of Consultant Committee on International Cooperation of National Natural Science Foundation of China. His research interests mainly focus on wireless communication. He is an Academician of the Chinese Academy of Engineering (CAE).
\end{CCJNLbiography}

\begin{CCJNLbiography}{photo_Dai.eps}{Jincheng Dai}
received the B.S. and Ph.D. degrees from Beijing University of Posts and Telecommunications (BUPT), Beijing, China, in 2014 and 2019. He is currently with the School of Artificial Intelligence, BUPT. His research interests include semantic communications, source and channel coding, machine learning for communications.
\end{CCJNLbiography}

\begin{CCJNLbiography}{photo_Si.eps}{Zhongwei Si}
received the Ph.D. degree from the KTH Royal Institute of Technology, Sweden, in 2013. She is currently an Associate Professor with the School of Artificial Intelligence, BUPT. Her research interests include wireless communication, information theory, and data mining.
\end{CCJNLbiography}

\begin{CCJNLbiography}{photo_Dong.eps}{Chao Dong}
received the B.S. and Ph.D. degrees from Beijing University of Posts and Telecommunications (BUPT), Beijing, China, in 2007 and 2012. He is currently an Associate Professor with the School of Artificial Intelligence, BUPT. His research interests include MIMO signal processing, multiuser precoding, decision feedback equalizer, and relay signal processing.
\end{CCJNLbiography}

\end{document}